\documentclass[final,5p,times]{elsarticle}

\usepackage{amssymb}
\usepackage{amsmath}

\usepackage[utf8]{inputenc}
\usepackage[table]{xcolor}
\usepackage{todonotes}
\usepackage{comment}
\usepackage{url}

\usepackage{xspace}
\usepackage{multirow}
\usepackage{float}
\usepackage{array}
\usepackage{tikz}
\usepackage{booktabs}

\usepackage{threeparttable}
\usepackage{textcomp}
\usepackage{overpic}
\usepackage{pgfplots}
\pgfplotsset{compat=newest}
\pgfplotsset{plot coordinates/math parser=false}
\usetikzlibrary{plotmarks}
\usetikzlibrary{mindmap,shadows,patterns,calc}
\usepackage[framemethod=tikz]{mdframed}
\definecolor{usethiscolorhere}{rgb}{0.86666,0.78431,0.78431}

\usepackage{amssymb}
\usepackage{booktabs}
\usepackage{pifont}
\usepackage{multirow}
\usepackage{paralist}

\usepackage{graphicx}
\usepackage{caption}
\usepackage{wrapfig}
\usepackage{xcolor}
\usepackage{subcaption}

\usepackage[ruled,norelsize]{algorithm2e}
\makeatletter
\newcommand{\removelatexerror}{\let\@latex@error\@gobble}
\makeatother

\usepackage{listings}
\newcommand\fix[1]{\textcolor{blue}{#1}}

\DeclareCaptionLabelFormat{alglabel}{Alg.~#2}

\usepackage[final,bookmarks=false]{hyperref}
\usepackage{worldflags}

\let\subparagraph\paragraph
\usepackage{titlesec}

\usepackage{cleveref}
\usepackage{caption}

\journal{Future Generation Computer Systems}

\begin{document}

\begin{frontmatter}

\title{
Enabling mixed-precision in spectral element codes}

\author[label1]{Yanxiang Chen} 
\author[label2]{Pablo de Oliveira Castro}
\author[label1]{Paolo Bientinesi}
\author[label4]{Niclas Jansson}
\author[label1,label3]{Roman Iakymchuk}
\affiliation[label1]{organization={Ume\aa\  University}, country={Sweden}}
\affiliation[label2]{organization={Université Paris-Saclay, UVSQ, LI-PaRAD}, country={France}}
\affiliation[label3]{organization={Uppsala University}, country={Sweden}} 
\affiliation[label4]{organization={KTH - Royal Institute of Technology}, country={Sweden}}

\begin{abstract}
Mixed-precision computing has the potential to significantly reduce the cost of exascale computations, but determining when and how to implement it in programs can be challenging.
In this article, we propose a methodology for enabling mixed-precision with the help of computer arithmetic tools, roofline model, and computer arithmetic techniques. 
As case studies, we consider Nekbone\fix{~\cite{nekbone}}, a mini-application for the Computational Fluid Dynamics (CFD) solver Nek5000\fix{~\cite{nek5000}}, and a modern Neko\fix{~\cite{neko}} CFD application. With the help of the Verificarlo\fix{~\cite{Denis2016verificarlo}} tool and computer arithmetic techniques, we introduce a strategy to address stagnation issues in the preconditioned Conjugate Gradient method in Nekbone and apply these insights to implement a mixed-precision version
of Neko. 
We evaluate the derived mixed-precision versions of these codes by combining metrics in three dimensions: accuracy, time-to-solution, and energy-to-solution. 
Notably, mixed-precision in Nekbone reduces time-to-solution by roughly 1.62x and energy-to-solution by 2.43x on MareNostrum 5, while in the real-world Neko application, the gain is up to 1.3x in both time and energy, with the accuracy that matches double-precision results.
\end{abstract}

\begin{keyword}
Mixed-precision \sep computer arithmetic tool \sep Verificarlo \sep roofline model \sep Conjugate Gradient \sep energy-to-solution \sep Neko.

\end{keyword}

\end{frontmatter}

\section{Introduction}

The energy consumption constraint for large-scale computing encourages scientists to revise the architecture design of hardware, linear algebra algorithms, and applications. The main idea is to make the computing cost sustainable and apply the \emph{lagom} principle (in Swedish: just the right amount), especially regarding working and storage precision. The gain in reducing and mixing precision brings not only faster time-to-solution but also a better energy footprint. Applications are relatively slow in picking the trend of energy-efficient computing due to their long-standing development (often over decades) and both complex and sophisticated codes with thousands, if not millions, of lines of code. 
Many applications share one thing in common: Most of their execution time is spent in a small set of computation kernels. 

In this article, we bridge the gap between algorithmic advances in mixed-precision numerical linear algebra and practical application development by leveraging state-of-the-art computer arithmetic tools.
This work extends our previous PPAM conference paper~\cite{ppam-paper} on mixed-precision in Nekbone by not only addressing stagnation issues and clarifying the critical role of precision in gather-scatter operations~\cite{Deville-Fischer-Mund-2002} and global communications but also by applying these insights to implement a mixed-precision version of Neko. Furthermore, we provide a detailed characterization of the energy and time-to-solution gains achieved on two European flagship supercomputing systems -- LUMI and MareNostrum 5 -- demonstrating significant improvements in both metrics.

Our main contributions are as follows:\\[-6mm]
\begin{enumerate}
\item We present a systematic methodology for enabling mixed-precision in CFD, specifically spectral element, codes. Using the computer arithmetic tool Verificarlo~\cite{Denis2016verificarlo}, we pinpoint key kernels for precision cropping. Verificarlo helps to identify precision issues but still requires expert validation. Verificarlo leverages Monte Carlo Arithmetic~\cite{parker97} and emulates variable precisions in order to simulate fluctuation in floating-point computations and evaluate the expected accuracy of the reduced precision computations.

\item Our analysis of the Nekbone~\cite{nekbone} mini-application shows that running the preconditioned Conjugate Gradient (CG)~\cite{Saa03} solver in single precision while retaining double precision for critical operations -- such as global communication and preconditioning -- can reduce time-to-solution by up to 1.62x and energy-to-solution by up to 2.43x on 80 MPI ranks, while resolving convergence stagnation issues.

\item We extend this mixed-precision strategy to the modern CFD application Neko. We demonstrate that using single precision for the solver with double precision for global reductions and key operations yields reductions of up to 1.3x in both execution time and energy consumption across two different preconditioners.

\item Extensive profiling, roofline modeling, and experiments on three HPC systems, including LUMI and MareNostrum 5, validate our methodology and highlight energy-to-solution as a key performance metric in exascale computing.

\end{enumerate}
\vspace{-2mm}

The rest of this article is organized as follows: \Cref{sec:background} describes floating-point arithmetic and Verificarlo. \Cref{sec:methodology} introduces our methodology that builds on three pillars: 1) code inspection with Verificarlo; 2) strategy to enable mixed-precision; 3) validation and verification in applications. As case studies, \Cref{sec:nekbone} covers Nekbone and \Cref{sec:neko} extends to Neko; in both, we inspect the codes and derive suitable mixed-precision strategies. \Cref{sec:results} discusses the main outcomes and benefits of enabling mixed-precision in both codes. Finally, \Cref{sec:conclusion} concludes the work and outlines future work.

\section{Background}
\label{sec:background}

In this section, we give a quick overview of two essential components in our methodology: Computer arithmetic, known as floating point arithmetic, and the software tool Verificarlo.

\subsection{Floating-Point Arithmetic} 

First, we briefly explain {\bf floating-point (FP) arithmetic} that
consists of approximating real numbers 
by numbers that have a finite fixed-precision representation adhering to the IEEE 754 standard. For instance, a FP number is represented on computers with a significand, an exponent, and a sign:
\vspace*{-3mm}
$$x = \pm \underbrace{x_0 . x_1 \ldots x_{M-1}}_{mantissa} \times\, b^{e}, \,\, 0 \leq x_i \leq b-1, \,\, x_0 \neq 0,$$\\[-4mm]
where $b$ is the  basis ($2$ in our case), $M$ is the precision, and $e$ stands for the exponent, i.e. range. 

The IEEE 754 standard~\cite{IEEE7542019}
specifies FP formats, which are often associated with precisions like {\em binary16} (also {\em fp16}), {\em binary32} ({\em fp32}), and {\em binary64} ({\em fp64}), see~\Cref{tb:ieee754}.
FP representation allows numbers to cover a wide 
\textit{dynamic range} that
is defined as the absolute ratio between the number with the largest magnitude and the number with the smallest non-zero magnitude in a set. All FP numbers are normalized and start with `1' as the first bit. It is also possible, but to a certain limited extend, to represent denormal numbers (denormals) that are any non-zero numbers whose magnitude is smaller than the smallest normalized floating-point number.

\begin{table}[ht]

\caption{Parameters for three IEEE arithmetic precisions.}
\label{tb:ieee754}
\centering
\begin{tabular}{lllll}
\hline
Type & Size & Signif. & Exp. & Rounding unit \\ 
\hline\noalign{\vskip .5mm} 
half & 16 bits & 11 bits & 5 bits & $u = 2^{-10} \approx 9.77 \times 10^{-4}$ \\ 
single & 32 bits & 24 bits & 8 bits & $u = 2^{-23} \approx 1.19 \times 10^{-7}$ \\ 
double & 64 bits & 53 bits & 11 bits & $u = 2^{-52} \approx 2.22 \times 10^{-16}$ \\ 
\hline
\end{tabular}
\end{table}

The standard requires correctly rounded results for the basic arithmetic operations $(+, -, \times , /, \sqrt{~},$ {\tt fma}$)$. 
This means that these operations are performed as if the results were first computed with infinite precision and then rounded to the FP format. The correct rounding rule guarantees a unique, deterministic, and well-defined answer.

\subsection{Verificarlo}
{\bf Verificarlo}~\cite{Denis2016verificarlo, verificarloproject} is an open-source tool, built upon the LLVM compiler, to analyze and optimize floating-point computations in large programs.   At compilation time, Verificarlo replaces each floating point operation by a custom call. At runtime, the program can be executed with various backends to explore different FP issues and optimisations. Verificarlo instruments FP operations at the optimised Intermediate Representation level (IR), yielding two main advantages: 1) Verificarlo can operate on any source language supported by the LLVM ecosystem, including C and C++ through {\tt clang}, and Fortran through {\tt flang}; 2) Verificarlo instrumentation takes place after all the other front-end and middle-end optimisation passes, thus capturing most compiler effects on FP operations. 

The two major backends are the Variable Precision (VPREC) backend and the Monte Carlo Arithmetic (MCA) backend. 

\textbf{VPREC}~\cite{Chatelain2019automatic} is a backend that transparently emulates lower precisions that fit into the original FP type. For example, if the original program uses {\tt binary64} (double) numbers, the user can emulate FP formats with a pseudo-mantissa of size $t \in [1, 52]$  and an exponent of size $r \in [2, 11]$.  VPREC intercepts each FP operation, performs the computation in double precision, and rounds the result to the emulated precision and range. VPREC has been carefully designed to correctly handle overflow, underflow, and denormals. 

\textbf{MCA} is a backend that implements different stochastic rounding modes to be used for estimating the effects and propagation of numerical errors in large programs.  MCA simulates the effects of different FP precisions by operating with a virtual precision $t$. To model errors on a value $x$, MCA uses the noise function $\text{inexact}(x) = x + 2^{e_x-t}\xi$,
where $e_x = \lfloor \log_2 \left|x\right| \rfloor +1 $ is the order of magnitude of $x$, and $\xi$ is a uniformly distributed random variable in the range $\left(-\frac{1}{2},\frac{1}{2}\right)$. When a program is run with the MCA backend, the result and/or the operands of each FP operation are replaced by a perturbed computation modeling the losses of accuracy~\cite{parker97}. 
In this study, we use two variants of MCA, which correspond to different substitutions of a FP operation \mbox{$y\circ z$} where $\circ \in \{+, -, *, / \}$: a) Random Rounding (RR) only introduces perturbation on the result: \(\text{round}(\text{inexact}(y\circ z))\); b) full MCA (MCA) introduces perturbation both on input operands and the result: \(\text{round}(\text{inexact}(\text{inexact}(y)\circ \text{inexact}(z)))\).

\section{Methodology}
\label{sec:methodology}
\begin{figure}[!ht]
\centering
\resizebox{\linewidth}{!}{\includegraphics{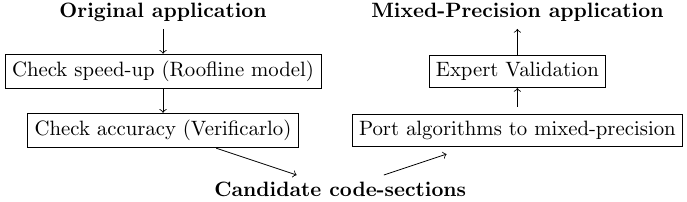}}
\caption{Methodology flow-chart. \label{fig:flow}}
\end{figure}

Adapting an application to use mixed-precision is a challenging and time-consuming process. It involves the manual modification of data-structures types, and of both mathematical and communication library calls.
It is still unclear if, and to what extent, it is possible to automate such a process. 
Here we propose a methodology, captured in~\Cref{fig:flow}, to assess which parts of a program can benefit from such work with the help of tools. Candidate code sections must both achieve a significant speed-up when using lower precision (eg. {\tt binary32}) data types and operations (\emph{speed-up check}) and achieve required accuracy with lower-precision operations (\emph{accuracy check}).

First, the program is profiled to identify the most-time consuming code sections and, possibly, produce roofline models, which we also use to evaluate the potential speed-up of moving computation and communications to lower precision.

Then, we check opportunities for precision cropping with the Verificarlo VPREC and MCA backends. The user must specify a target accuracy threshold on one or multiple outputs; optimizations that degrade accuracy over the target threshold will not be considered. This threshold is application and quantity dependent; usually, it is chosen by the application owner who knows how many significant digits are needed for a given computation.
The backends estimate the error introduced by lower, eg. {\tt binary32}, precision on the designated code sections. VPREC is used first on each section separately. The forward error of the output is monitored. If it raises above a user's defined threshold, the section is filtered out. The set of candidate sections is further pruned thanks to the MCA backend, which eliminates sections that are unstable with stochastic rounding. 

Regarding the accuracy, this methodology does not give formal guarantees on the error for arbitrary datasets. However, by using different datasets in the pruning step, the user can efficiently eliminate code sections that are too sensitive to lower precision computations. 
Our contribution consists, therefore, in providing a step-by-step tool-driven 
methodology for reducing the scope of code considered for porting to lower precision.

After the automatic pruning, the expert should modify the code to use lower precision when needed and validate the new algorithm error guarantees. This step follows with time-to-solution and energy-to-solution measurements.
In the following sections, we demonstrate this methodology on the Nekbone mini-app as well as on the Neko application.

\section{Nekbone case study}
\label{sec:nekbone}
Nekbone is a mini-app that captures the basic structure and design of the extensive Nek5000~\cite{nek5000} software, which is a high-order, incompressible Navier-Stokes solver based on the spectral element method. Nekbone solves a standard Poisson equation by partitioning the computational domain into high-order quadrilateral elements, and by using the Conjugate Gradient (CG) method with a multigrid preconditioner (the preconditioner is optional and can be enabled when compiling). 

The main computational kernel is the Conjugate Gradient loop. 
Overall, each CG iteration consists of vector operations, matrix-matrix multiply operations, nearest-neighbor communications with MPI, and MPI Allreduce operations.
Nekbone’s implementation organizes the elemental operations tensor products via the dense matrix–matrix multiplications of relatively small sizes. In spectral element methods, each element carries its own small mass or stiffness matrix, and applying these local operators amounts to performing matrix–matrix multiplies on element‐local blocks.

The main computational kernel (the CG loop) in Nekbone is written in Fortran 77, and the Gather-Scatter library~\cite{gslib} (an individual component of Nek5000) for the nearest-neighbor communications is written in C plus MPI. 
Although both double and single precision data types are technically compatible with some of Nekbone's components, Nekbone is developed entirely in double precision. Thus, Nekbone cannot be run completely in single and mixed single plus double precision was not envisioned by the original design.

\subsection{Code inspection with tools}
\label{sec:code-inspection}
We use a few tools in order to identify computationally intensive parts of the code, to have a projection on the expected performance, but primarily to inspect the code for numerical abnormalities and possibilities for precision reduction.

\Cref{tab:cg-profiling} shows the code for the CG loop, the operations associated with each line, and the profiling results, which consist of include time (just {\em time} for simplicity). Time is the percentage of the overall runtime. There is also a possibility to retrieve self time, which is the percentage of the runtime. If a kernel has no callees, then the time equals the self time. Since this is often the case, we only report the time.

\begin{figure}
    \centering
    \resizebox{0.9\linewidth}{!}{\includegraphics{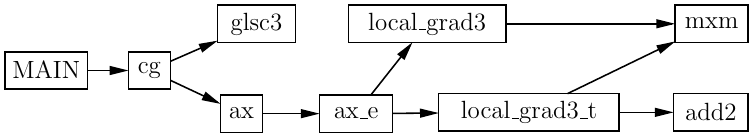}}
    \caption{Call graph of Nekbone without preconditioner, only the most time-consuming computational kernels are shown.}
    \label{fig:callgraph}
\end{figure}

\begin{figure}
    \centering
    \resizebox{1.05\linewidth}{!}{\includegraphics{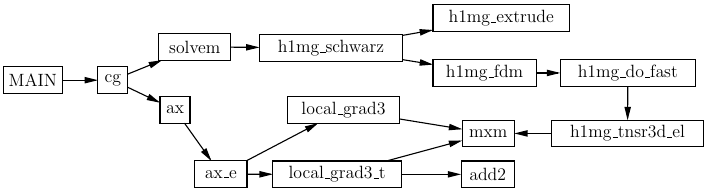}}
    \caption{Call graph of Nekbone with the preconditioner enabled.}
    \label{fig:callgraph-precond}
\end{figure}

The profiling results are obtained by the {\em Callgrind tool}~\cite{callgrind} tool 
(a module of Valgrind~\cite{valgrind}); the column no-MGRID shows the timings without the multigrid preconditioner. The most time-con\-suming kernel is the sparse matrix-vector multiplication ({\tt ax}), which calls the {\tt local\_grad3}, {\tt local\_grad3\_t}, {\tt add2} and {\tt mxm} (the call graph is shown in~\Cref{fig:callgraph}). Conversely, when the preconditioner is enabled, the application of the preconditioner ({\tt solveM}) is the most time-consuming, followed closely by {\tt ax}
(the call graph is shown in~\Cref{fig:callgraph-precond}).
\begin{table*}
\caption{Profiling of the CG kernels within Nekbone.} 

\label{tab:cg-profiling}
\centering
\begin{threeparttable}[t]
\begin{tabular}{p{0.26\textwidth}|p{0.127\textwidth}p{0.09\textwidth}p{0.09\textwidth}}
\hline \noalign{\hrule height 0.3pt}
 CG loop\tnote{a} & Description & \multicolumn{2}{c}{Time (\%)} \\
\hline
1 \ \textbf{do} iter = 1, miter &  & no MGRID & MGRID \\
\cline{3-4} 
2 \ \ \ \ call \textbf{solveM}(z,r,n) & preconditioner & - & 69.82 \\
3 \ \ \ \ $\beta$ = \textbf{glsc3}(r,c,z,n) & dot product & 5.79\tnote{b} & 1.72 \\
4 \ \ \ \ call \textbf{add2s1}(p,z,$\beta$,n) & vector add & 1.76 & 0.52 \\
5 \ \ \ \ call \textbf{ax}(w,p,g,ur,us,ut,wk,n) & multiplication\tnote{c} & 86.73 & 25.83 \\
6 \ \ \ \ pap = \textbf{glsc3}(w,c,p,n) & dot product & 5.79 & 1.72 \\
7 \ \ \ \ call \textbf{add2s2}(x,p,pap,n) & vector add & 5.05 & 1.50 \\
8 \ \ \ \ call \textbf{add2s2}(r,w,pap,n) & vector add & 5.05 & 1.50 \\
9 \ \ \ \ rtr = \textbf{glsc3}(r,c,r,n) & dot product & 5.79 & 1.72 \\
10 \ \textbf{enddo} &  & & \\
\hline \noalign{\hrule height 0.3pt}
\end{tabular}
\begin{tablenotes}
    \item[a] We include only those lines that contain compute kernels. 
    \item[b] Multiple calls to the same kernel are accumulated, e.g. lines 3,6,9.    
    \item[c] Matrix-matrix multiplications, and accumulations before and after.
\end{tablenotes}
\end{threeparttable}
\end{table*}

A thorough inspection of the application helps to identify computational bottlenecks and to assess where mixed-precision can be safely applied. We use Verificarlo to analyze numerical stability and identify sections where reducing precision does not compromise accuracy. The inspection focuses on the following two steps:

\begin{enumerate}
    \item Simulating the application using the VPREC backend with lower precisions and assessing the potential for precision reduction by tracking a few significant variables such as residual, $\beta$, {\tt pap}, and {\tt rtr} in the (preconditioned) CG algorithm. Computation of dot products, as for the residual ({\tt rtr}), require parallel reductions that are the main source for error accumulation and, hence, inaccuracy in iterative solvers as shown in~\cite{iakymchuk19jcam}.
    \item Further verifying convergence sensitivity using stochastic arithmetic through the MCA backend.
\end{enumerate}

We perform the analysis with Verificarlo version 1.0, Flang and LLVM version 14.0.1 (Classic Flang Project) 
installed on Kebnekaise at HPC2N at Ume\aa{} University. We present more details of Kebnekaise in~\Cref{sec:results}.

\subsubsection{Lower precision emulations with Verificarlo VPREC}

\begin{figure*}[!t]
    \hspace{-2mm}
    \begin{subfigure}{0.51\textwidth}
        \centering
        \resizebox{0.7\linewidth}{!}{\includegraphics{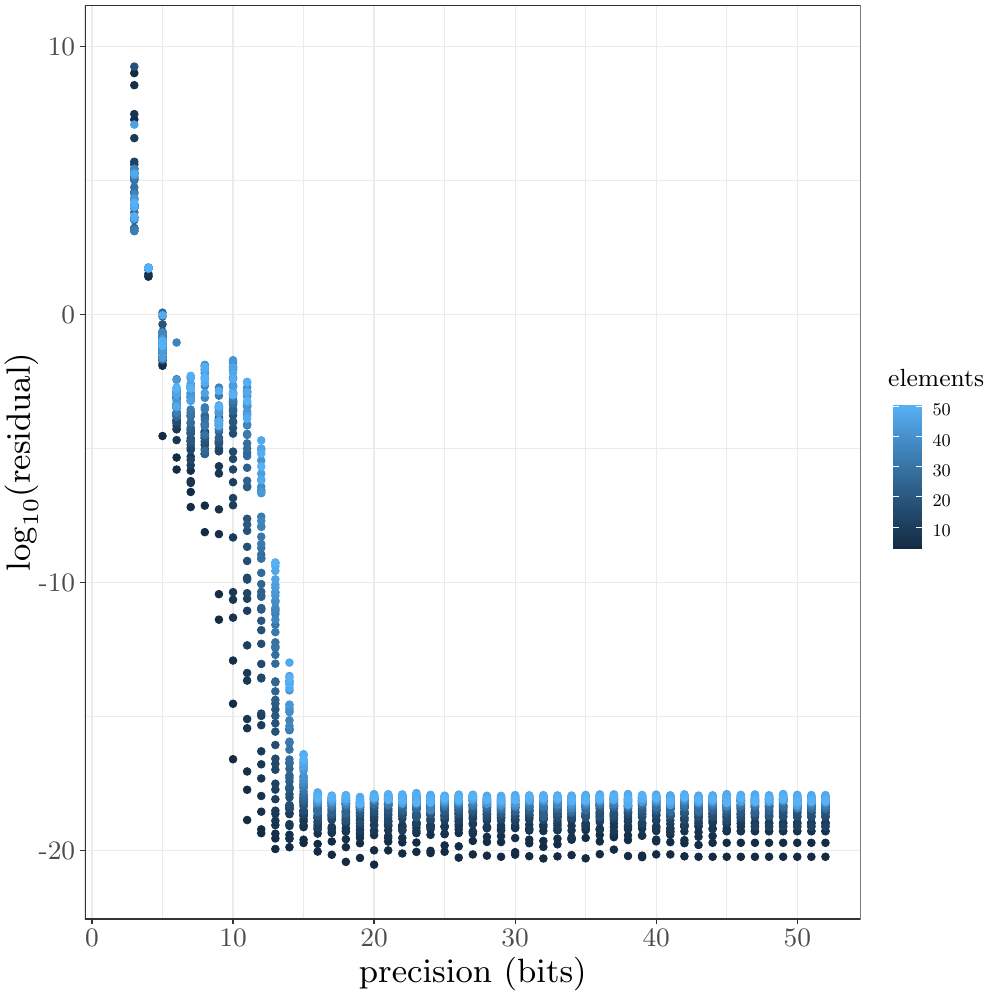}}
        \caption{Whole program} \label{fig:vprec-whole}
    \end{subfigure}
    \hspace{0.7mm}
    \begin{subfigure}{0.51\textwidth}
        \centering
        \resizebox{0.7\linewidth}{!}{\includegraphics{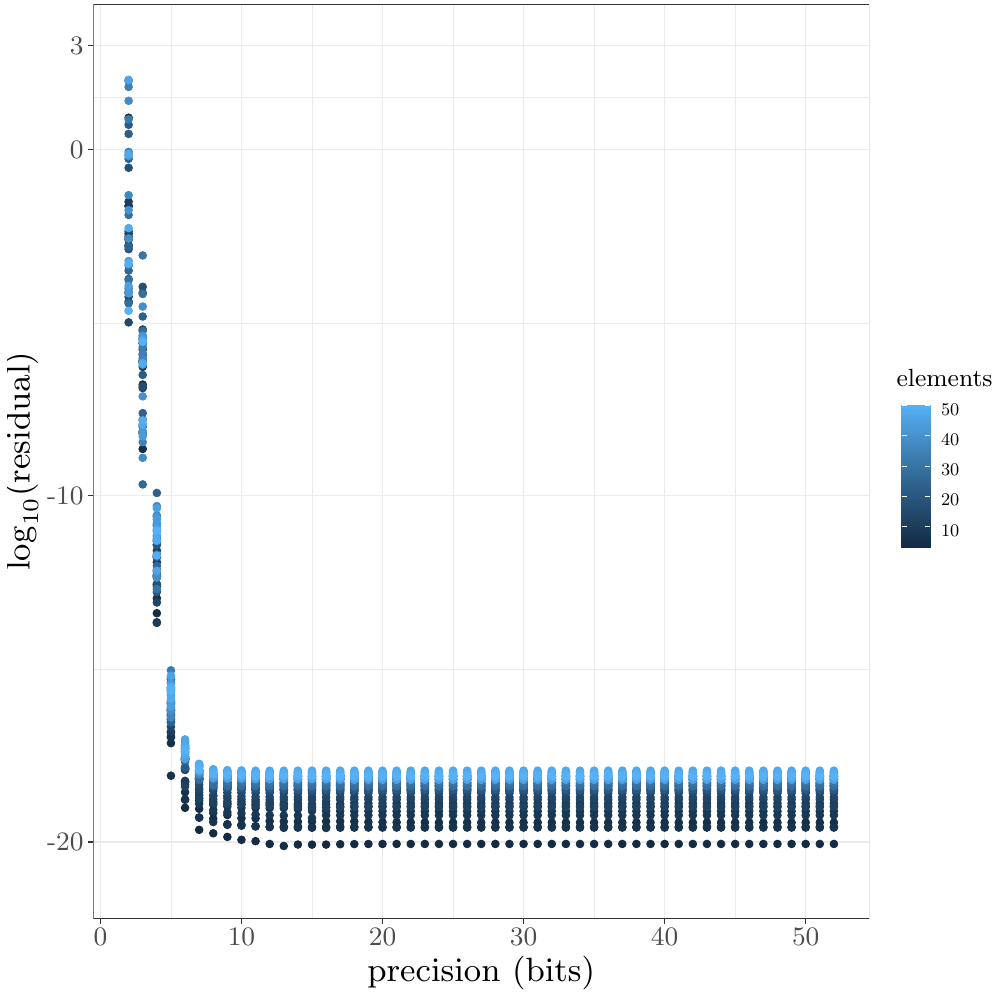}}
        \caption{Only CG} \label{fig:vprec-cg}
    \end{subfigure}
    \caption{Emulating different precision (mantissa) from 3 to 52 bits with the VPREC backend in Nekbone for different spectral elements and observing the change in the residual in CG without preconditioner.}
    \label{fig:vprec-whole-cg}
\end{figure*}

We use the VPREC backend from Verificarlo to instrument double precision FP operations, simulating lower precisions with $t\in [3,52]$ for the whole Nekbone application; we also vary the number of spectral elements, where higher is better, reassembling more realistic simulations. 
Note that we change only the mantissa while keeping the exponent as in higher precision, double in this case. 
\Cref{fig:vprec-whole} illustrates how $residual$ becomes smaller as precision grows, stabilising at $t \geq 16$ at its lowest level. We track the residual ({\tt rtr}) because it captures the overall effect of lowering precision on the solution accuracy. Both $\beta$ and {\tt pap} follow the trend.

When only the functions involved in the CG kernel are instrumented, the plot becomes even smoother and reaches a stable state at $t \geq 8$, as depicted in~\Cref{fig:vprec-cg}. The results of the VPREC experiments indicate that there is a potential for single precision in the entire program, leaving enough space for the error accumulation. After this initial evaluation, we test the sensitivity of these results with the MCA backend.

\subsubsection{Sensitivity analysis with Verificarlo MCA}
\begin{figure}[!t]
    \begin{subfigure}{0.49\textwidth}
        \centering
        \includegraphics[width=0.8\linewidth]{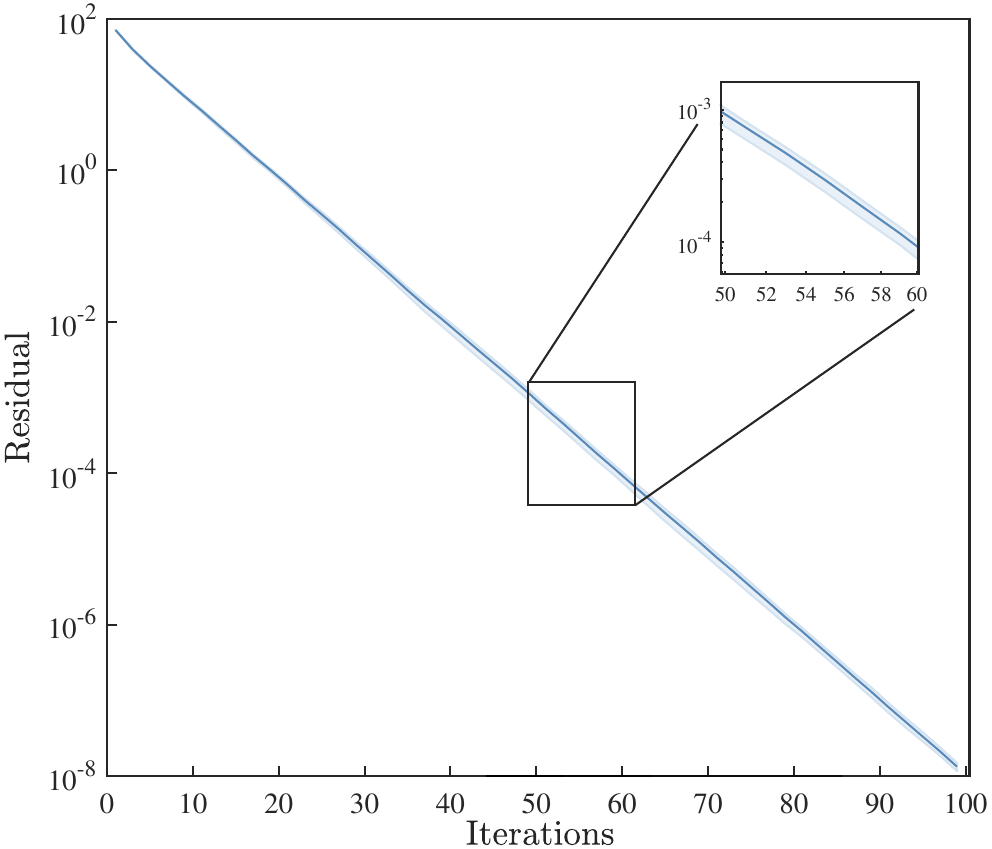}
        \caption{$rr$ mode} \label{fig:mca-rr}
    \end{subfigure}
    \hspace*{\fill}
    \begin{subfigure}{0.49\textwidth}
        \centering
        \includegraphics[width=0.8\linewidth]{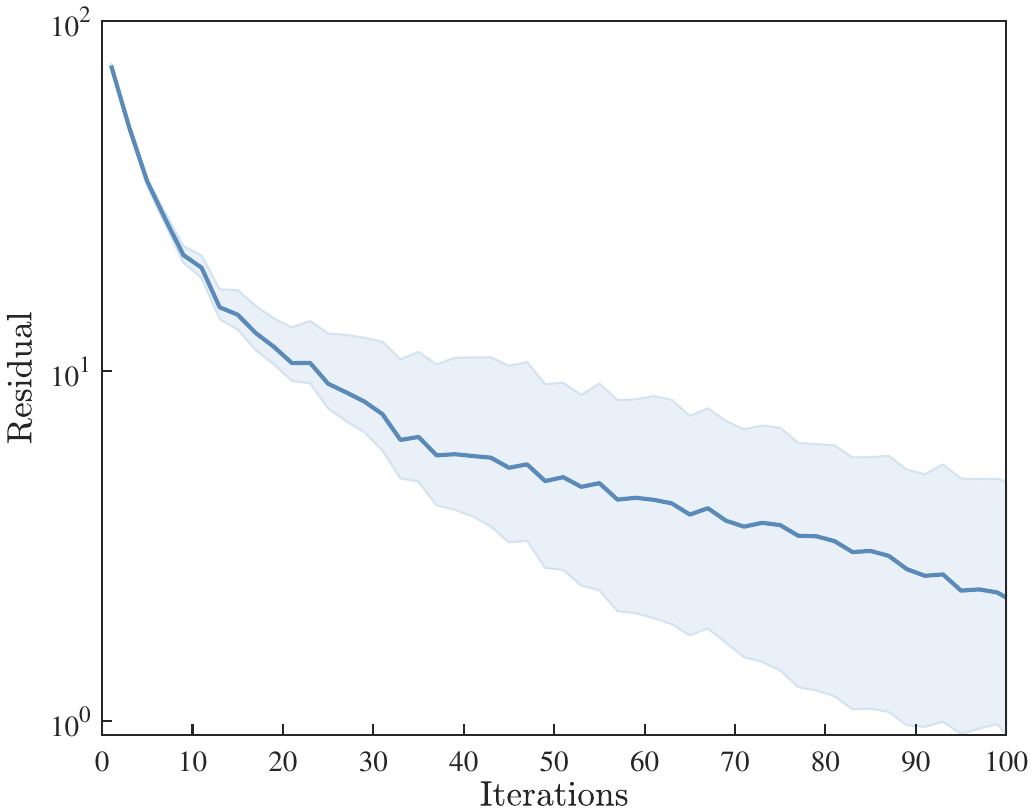}
        \caption{$mca$ mode} \label{fig:mca-mca}
    \end{subfigure}    
    \caption{Residual history of CG: The sensitivity inspection with the MCA backend in the {\em entire Nekbone}.} 
    \label{fig:mca-whole}    
\end{figure}

We first run the whole program using the MCA backend. The $mca$ and $rr$ modes are employed with the precision (mantissa) $t = 23$ bits. The exponent is kept as in double precision, so we emulate single precision within the double precision format to replace doubles. 
We run the code $20$ times with the MCA backend in order to have a statistically significant sample size. \Cref{fig:mca-whole} illustrates the error bar plot (filled area) for $20$ runs, the blue curve represents the mean value $\mu =\sum residual/ 20$ for each iteration, while the upper and lower bounds of the error bar are determined by the maximum and minimum values: $err=[ min,\ max]$.  
\begin{figure}[!t]
    \begin{subfigure}{0.49\textwidth}
        \centering
        \includegraphics[width=0.8\linewidth]{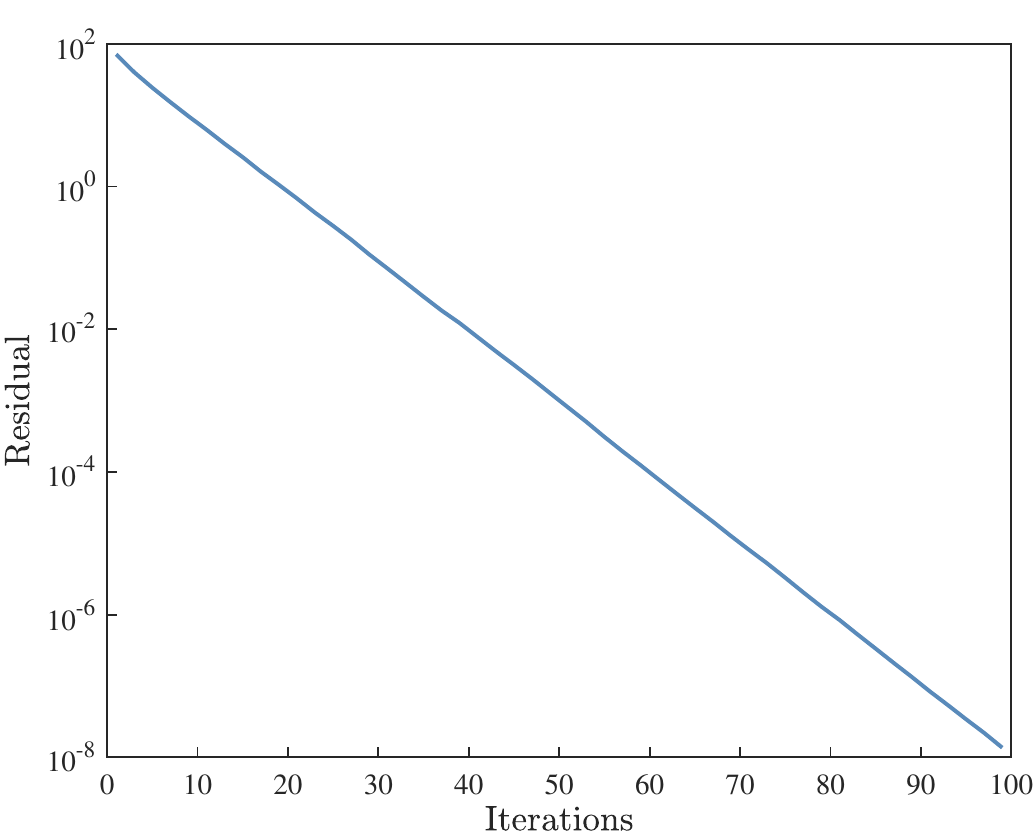}
        \caption{$rr$ mode} \label{fig:mca-rr-onlycg}
    \end{subfigure}
    \hspace*{\fill}
    \begin{subfigure}{0.49\textwidth}
        \centering
        \includegraphics[width=0.8\linewidth]{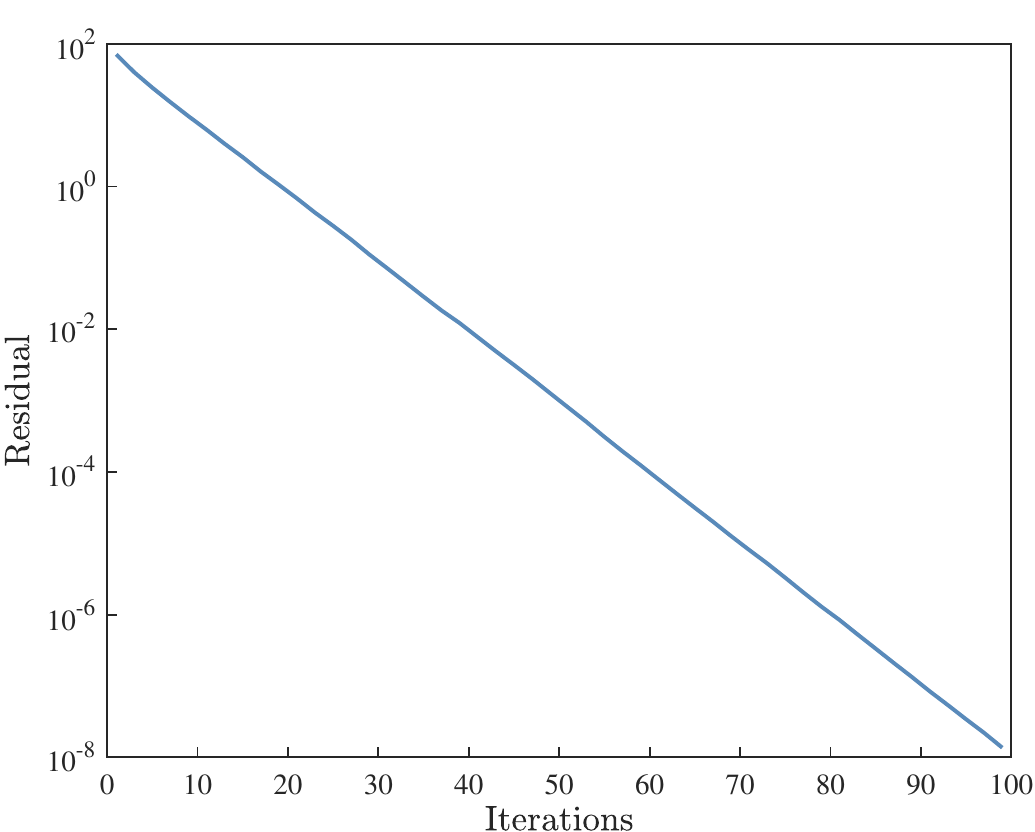}
        \caption{$mca$ mode} \label{fig:mca-mca-onlycg}
    \end{subfigure}
    \caption{
    Residual history of CG: The sensitivity inspection with the MCA backend in the {\em CG loop only} in Nekbone; .} 
    \label{fig:mca-onlycg} 
\end{figure}

In our analysis using the MCA backend, we observed negligible error in Random Rounding ($rr$) mode, indicating that the single-precision simulation generally yields acceptable results. However, significant error fluctuations occur when operating in the full MCA mode, eventually breaking convergence. In order to identify sources of the fluctuations, we subsequently conducted a thorough analysis of each subroutine separately using the $mca$ mode. 

We found that the function {\tt pnormj}, which is called during Nekbone’s initialization phase, is particularly sensitive to the precision reduction. Removing {\tt pnormj} from the MCA instrumentation reduced the fluctuations but did not eliminate them entirely. Only when all initialization routines were excluded did the fluctuations disappear. This suggests that certain initialization operations -- especially expressions like $10^9 * \cos(x)\in [0,10^9]$ -- are extremely sensitive since the span of the values is larger than what the single-precision mantissa can cover.

Considering the sensitivity of the initialization routines and the profiling results that identify the CG loop as the most time-consuming kernel, we decided to focus exclusively on the CG loop to explore precision cropping. \Cref{fig:mca-onlycg} depicts the results of our inspection with the $mca$ and $rr$ modes in the CG loop. These results show no fluctuations. 

Thus, the code inspection with both VPREC and MCA backends reveals a possibility for precision cropping from double to single in the CG loop without preconditioner. In particular, the MCA backend with both modes, introducing noise as an input and output in FP operations, tests the numerical sensitivity of such cropping and confirms this possibility. 
These encouraging initial results motivate us to investigate more general problem settings, including different boundary and initial conditions, different domains, and different meshes. By following this line, we also aim to include adaptivity in the proposed mixed-precision strategy enabling algorithms to adjust to the above-mentioned simulation changes.

\subsubsection{Roofline modeling}
\label{sec:roofline}

The roofline model~\cite{roofline} provides us with valuable insights for assessing and enhancing software for FP computations. This paper aims to identify the bottlenecks of the double precision Nekbone and evaluate the improvements achieved through mixed-precision using the roofline model for analysis. The model is generated with the help of Intel\textsuperscript{\textregistered} Advisor~\cite{advisor} version 2023.2.0 installed on Kebnekaise. We use Intel Compiler Toolkits version 2021.9.0, specifically {\tt ifort} and {\tt icc} compiler, with the {\tt -O2} optimization flag for the code compilation. Note that many optimizations including vectorization can be enabled with {\tt -O2}.

\begin{figure}[!t]
    \centering
    \includegraphics[width=\linewidth]{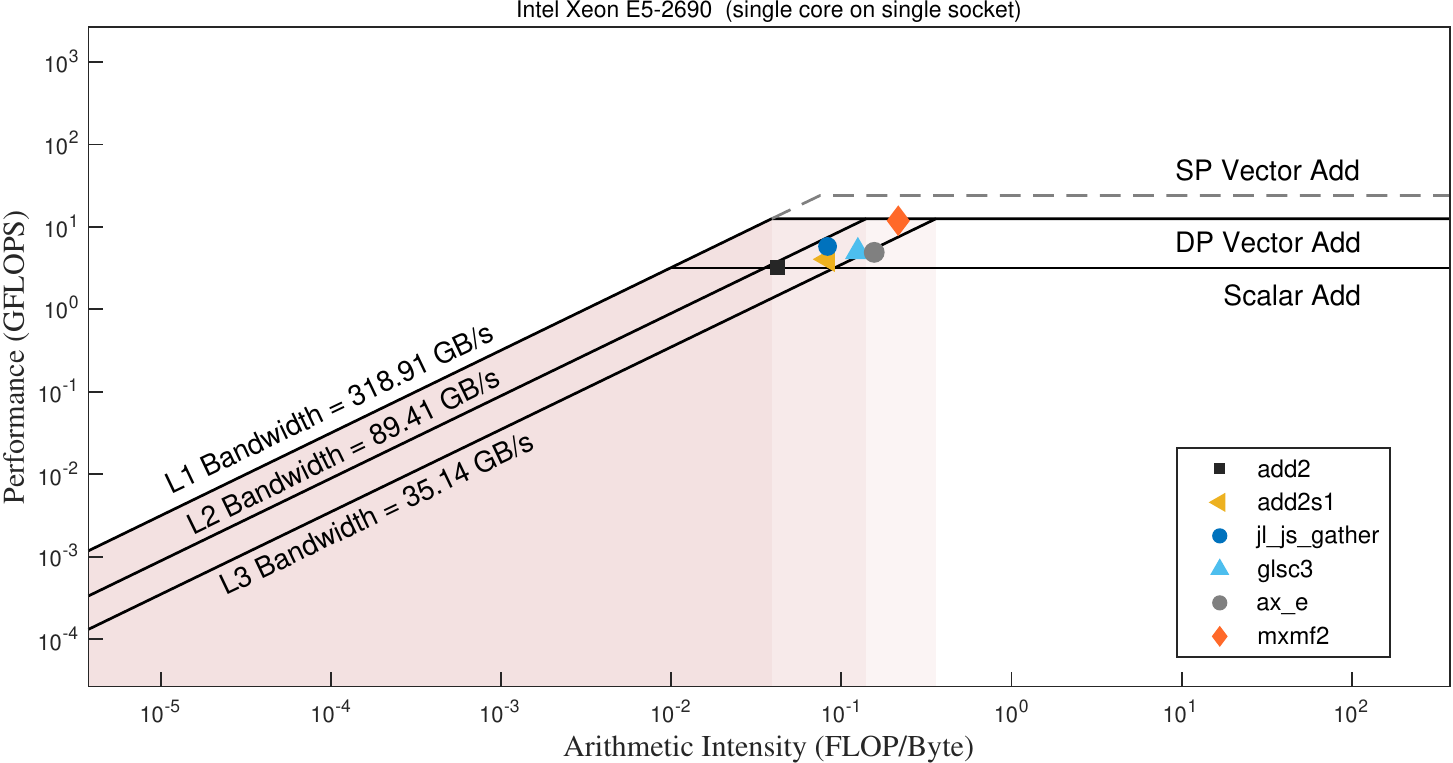}
    \caption{Roofline model for the original double precision Nekbone.}
    \label{fig:roofline-double}
\end{figure}

\Cref{fig:roofline-double} shows the roofline model for the case without preconditioner in double precision on an Intel Xeon E5-2690 processor (single core), with the elements from 1 to 100. The graph is plotted on a logarithmic scale, with three parallel diagonal lines representing the memory bounds Level (L1), L2, and L3 caches. The x-axis represents the arithmetic/ operational intensity (FLOP/ Byte), while the y-axis is the attainable floating-point performance, measured in GFLOPS. Horizontal lines on the graph represent different compute bounds:
\begin{itemize}
    \item SP Vector Add: single precision vector add peak bound (24.06 GFLOPS);
    \item DP Vector Add: double precision vector add peak bound (12.58 GFLOPS);
    \item Scalar Add: scalar add peak bound (3.17 GFLOPS).
\end{itemize}
The data points in the figure represent the computational kernels: {\tt add2}, {\tt add2s1}, {\tt glsc3}, and {\tt ax\_e} are limited by the L2 memory bound, while the {\tt mxmf2} (the implementation of matrix multiplication, called by {\tt mxm}) is limited by the DP Vector Add bound.

\begin{figure}[!ht]
    \centering
    \includegraphics[width=\linewidth]{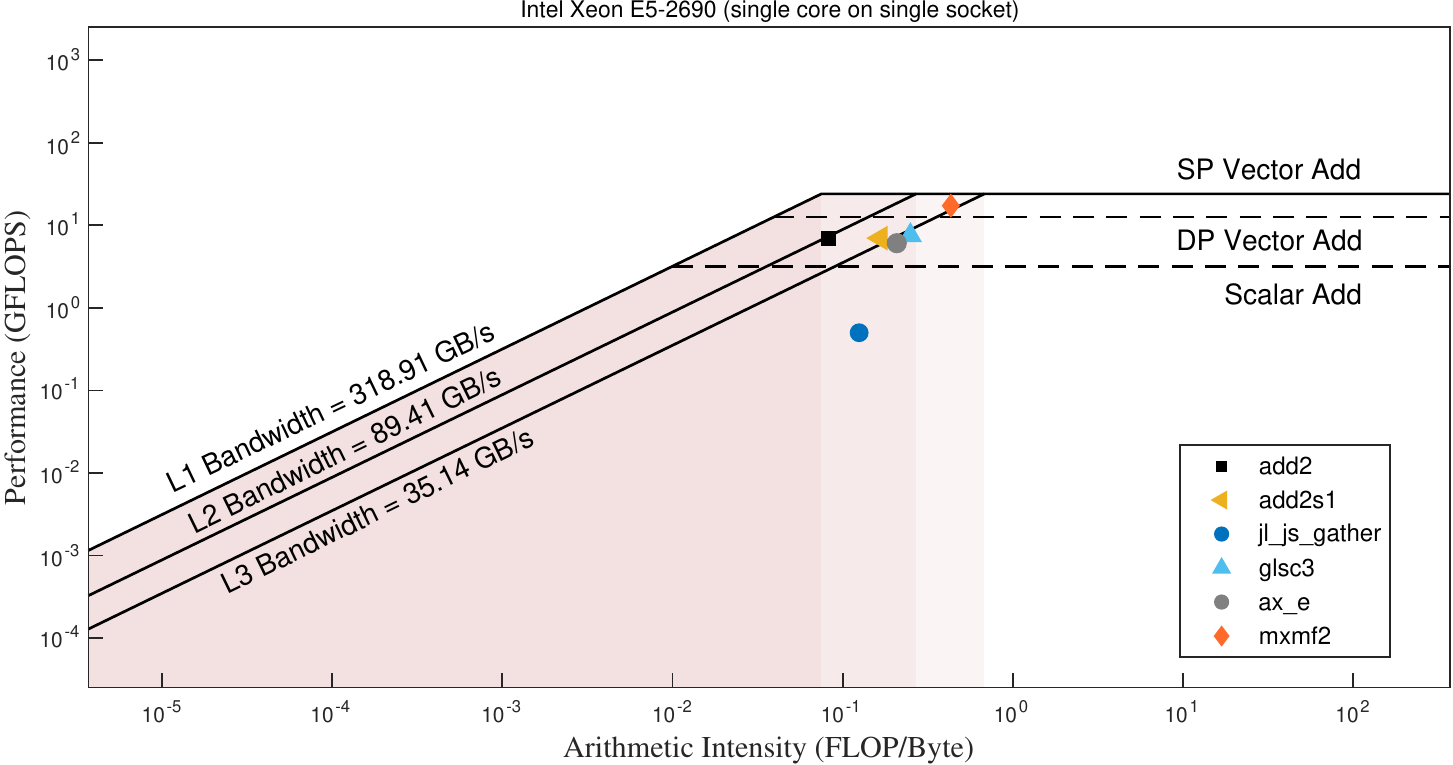}
    \caption{Roofline model for the mixed-precision Nekbone.}
    \label{fig:roofline-single}
\end{figure}

The implementation of the mixed-precision Nekbone, the same test case, has led to a notable improvement, see~\Cref{fig:roofline-single}:
{\tt mxmf2} surpassed the limits of DP Vector Add bound, with the remaining kernels break the L2 memory bound. \Cref{tab:roofline-dp-sp} presents a comparison between the original and mixed-precision versions.
\begin{table}[!ht]
\caption{Roofline bounds for the double and mixed-precision Nekbone.}
\label{tab:roofline-dp-sp}
\centering
\begin{tabular}{p{0.224\linewidth}|p{0.345\linewidth}p{0.338\linewidth}}
\hline \noalign{\hrule height 0.3pt}
Kernels & Double Precision & Mixed-precision \\
\hline 
{\tt add2} & L2 memory bound & SP Vector Add bound\\
{\tt add2s1} & L2 memory bound & L3 memory bound \\
{\tt glsc3} & L2 memory bound & L3 memory bound \\
{\tt ax\_e} & L2 memory bound & L3 memory bound \\
{\tt jl\_gs\_gather} & L1 memory bound & Scalar Add bound \\
{\tt mxmf2} & DP Vector Add bound & SP Vector Add bound \\
\hline \noalign{\hrule height 0.3pt}
\end{tabular}
\end{table}

\subsection{Inspection of the preconditioned CG}
\label{sec:precond-nekbone}
\begin{figure}[!ht]
    \centering

    \includegraphics[width=0.8\linewidth]{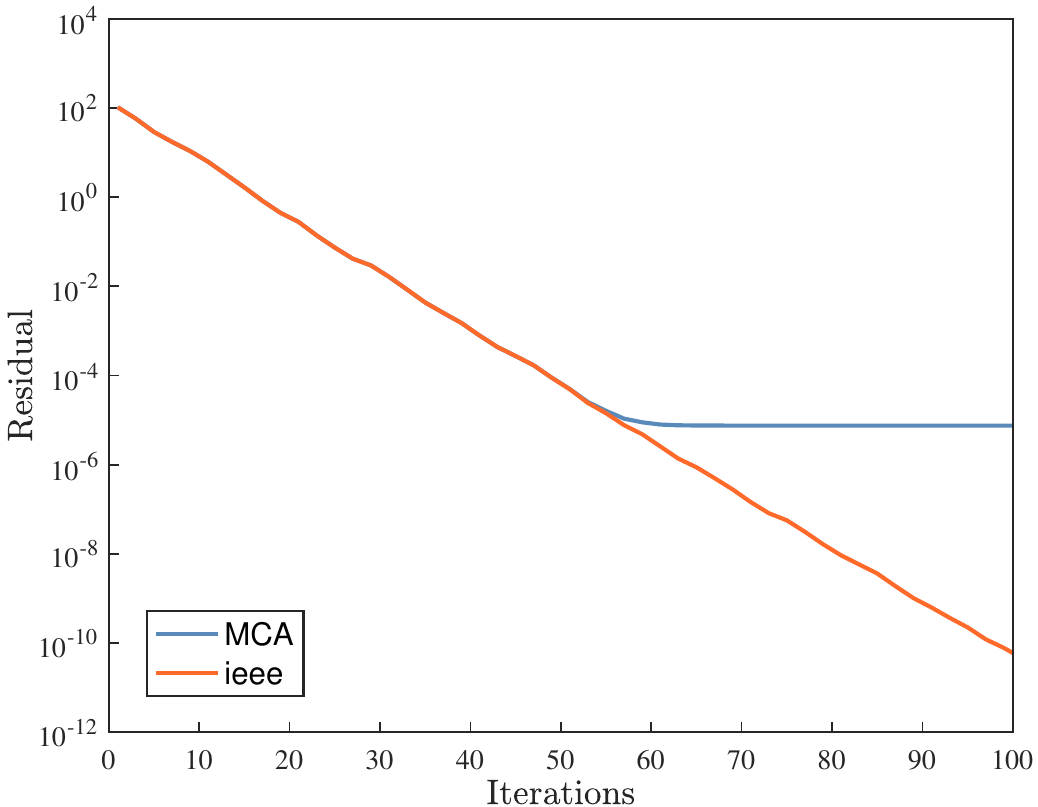}
    \caption{Mixed-precision Nekbone with the multigrid preconditioned CG inspected under the $mca$ mode in the MCA backend; $ieee$ stands for double precision results.}
    \label{fig:mca-precond}   
\end{figure}
We also inspect Nekbone with the multigrid preconditioned CG (PCG). We notice stagnation in both the $rr$ and $mca$ modes using MCA backend: As depicted in~\Cref{fig:mca-precond}, the residuals flattens after the $61$st iteration (with the residual of $7.94\times 10^{-6}$). These results demonstrate that static precision cropping is not applicable with the preconditioner enabled. 

The pinpointed routine \texttt{h1mg\_schwarz\_wt3d2} relies on square-root operations. We hypothesized that performing these operations in single precision might cause the observed stagnation. To test this hypothesis, we maintained the routine in fp32 but replaced the square-root operations with calls to \texttt{dqsrt} to perform them in fp64\footnote{We also tested the \texttt{-no-prec-sqrt} compiler option with the Intel Compiler, but it had no effect on convergence.}. This modification resolved the stagnation issue for cases with low parallelism (fewer than four MPI ranks). Despite the additional overhead from \texttt{dqsrt} calls and type casting, this version still achieved significant speedup (1.63x speedup, for serial test with 128 elements) compared to the original fp64 implementation.

\begin{figure*}
    \centering
    \includegraphics[width=\linewidth]{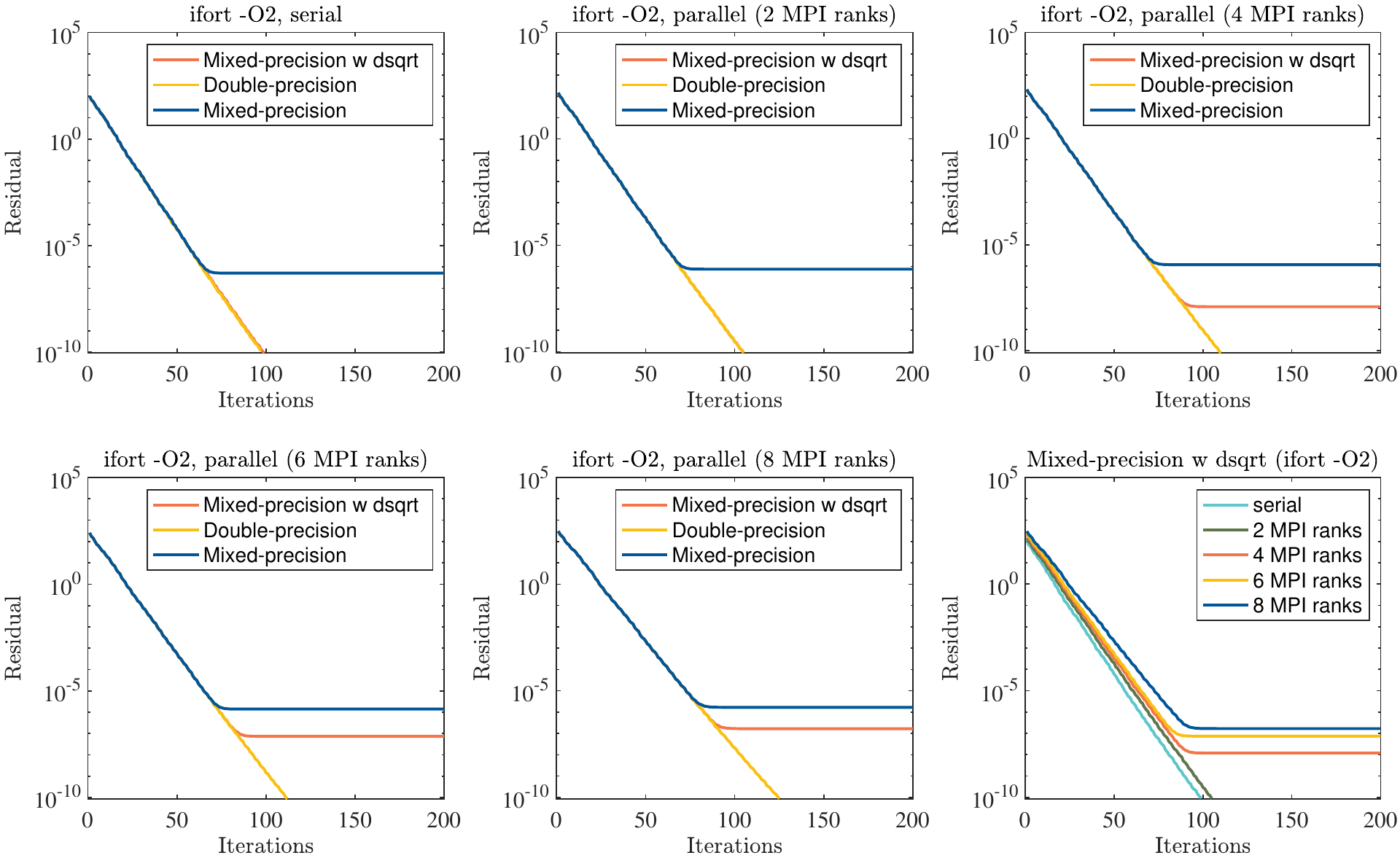}
    \caption{{\tt dsqrt} as the source of inaccuracy in the preconditioned CG in Nekbone.}
    \label{fig:dsqrt-serial-parallel}
\end{figure*}

However, in parallel configurations with more than four MPI ranks, stagnation persisted, though it occurred later in the computation (see \Cref{fig:dsqrt-serial-parallel}). The continued presence of stagnation in the parallel case suggests that the root cause lies in the global communication operations, which we discuss next. 

\subsubsection{Global communication and gather-scatter modes}

In spectral element codes like Nek5000/ Nekbone, computation proceeds entirely in a matrix-free method, that is, the global stiffness matrix is never assembled. Instead, continuity across elements is enforced by the combined gather-scatter operator $\Sigma':= Q\,Q^{T}$, often referred to as direct stiffness summation~\cite{Deville-Fischer-Mund-2002}. $Q$ is a scatter operator which maps the global vector $u$ to local elemental vectors $u_L$, followed by a gather operator $Q^T$, which sums entries from nodes to $u$. In Nek5000/ Nekbone, the gather-scatter is implemented by the gslib library~\cite{gslib}. In Neko, they have developed the overlapping gather-scatter kernel~\cite{neko}. Throughout this paper, we denote these operations as Gather-Scatter operations (GSO).

We investigated the impact of GSO, which perform reductions through either pairwise communication or standard allreduce operations (see \Cref{tab:evaluation}).  Initially, we test implementing these global communications in double precision separately in the solver and preconditioner, as well as in both components simultaneously. When double precision GSO is used exclusively in either the solver or the preconditioner while keeping other operations in single precision, the  convergence stagnated with four or more MPI ranks. However, when the double precision GSO were implemented in both the solver and preconditioner, parallel runs showed no stagnation. This observation leads to an important conclusion: Within the PCG loop, all global operations -- both in the solver and preconditioner -- must be computed in double precision, even when the local arithmetic operations are in the lower precision.

An alternative to increase precision in the GSO communication is to change the gather-scatter communication mode itself. Nekbone's gather-scatter library supports three modes, namely pairwise, crystal router, and allreduce, with the mode selection typically determined through automatic benchmarking. On Kebnekaise, the pairwise mode is tested to be the default. The allreduce mode has its potential depending on the architecture. Hence, we test these modes with single-precision communications. With eight MPI ranks and 128 elements per rank, the pairwise mode stagnated at $2.0\times 10^{-6}$, while both crystal router and allreduce modes converged to $10^{-10}$ after 124 and 134 iterations, respectively. However, at higher parallelism (28 ranks) with the same elements per rank, only the allreduce mode maintains convergence to $10^{-10}$ after 138 iterations, while both the pairwise and crystal router modes stagnate. These results suggest that the choice of communication mode can significantly impact convergence behavior in the mixed-precision implementations, with allreduce showing the most robust performance at higher levels of parallelism. We can further reinforce this mode for global communication with computer arithmetic techniques, which we discuss in~\Cref{sec:neko}.

To sum up, \Cref{tab:evaluation} presents various mixed-precision strategies, where the focus is on the ones that lead to the convergence. While the strategy with double precision preconditioner, see the last row of the upper table, converges,  it is worth noting that the benefit of mixing precisions will be demolished due to 70\,\% of the execution time spent in the preconditioner, see~\Cref{tab:cg-profiling}. From the lower table, the allreduce GSO mode clearly has its gain, but it's known to be architecture-specify, e.g. IBM Blue Gene, and also is not in use in Neko.

\begin{table}[!t]
\centering
\small
\begin{tabular}{@{}l@{\hspace{4pt}}c@{\hspace{4pt}}cc@{\hspace{4pt}}cc@{\hspace{4pt}}c@{}}
\toprule
\multicolumn{7}{c}{\textbf{Mixed-precision strategy evaluation}} \\
\midrule
\textbf{GSO mode} & \textbf{MPI} & \multicolumn{2}{c}{\textbf{Precond.}} & \multicolumn{2}{c}{\textbf{PCG}} & \textbf{Conv.} \\
\cmidrule(lr){3-4} \cmidrule(lr){5-6}
& & \textbf{Ops} & \textbf{GSO} & \textbf{Ops} & \textbf{GSO} & \\
\midrule
\multirow{9}{*}{pairwise} & \multirow{4}{*}{$<$ 4} & fp32 & fp32 & \multirow{9}{*}{fp32} & \multirow{4}{*}{fp32} & {\color{red}stagnates} \\
& & fp64 & fp64 & & & {\color{blue}converges} \\
& & fp32+fp64sqrt & fp32 & & & {\color{blue}converges} \\
& & fp32 & fp64 & & & {\color{blue}converges} \\
\cmidrule(lr){2-4} \cmidrule(lr){6-7}
& \multirow{5}{*}{$\geq$ 4} & fp64 & fp64 & & \multirow{3}{*}{fp32} & {\color{red}stagnates} \\
& & fp32+fp64sqrt & fp32 & & & {\color{red}stagnates} \\
& & fp32 & fp64 & & & {\color{red}stagnates} \\
& & fp32+fp64sqrt & fp64 & & fp64 & {\color{blue}converges} \\
& & fp64 & fp64 & & fp64 & {\color{blue}converges} \\
\midrule
\multicolumn{7}{c}{\textbf{Gather-Scatter mode evaluation}} \\
\midrule
\textbf{GSO mode} & \textbf{MPI} & \multicolumn{2}{c}{\textbf{Precond.}} & \multicolumn{2}{c}{\textbf{PCG}} & \textbf{Conv.} \\
\cmidrule(lr){3-4} \cmidrule(lr){5-6}
& & \textbf{Ops} & \textbf{GSO} & \textbf{Ops} & \textbf{GSO} & \\
\midrule
\multirow{2}{*}{pairwise} & 8 & \multirow{4}{*}[-\dimexpr \aboverulesep + \belowrulesep + \cmidrulewidth]{fp32} & \multirow{6}{*}[-\dimexpr \aboverulesep + \belowrulesep + \cmidrulewidth]{fp32} & \multirow{6}{*}[-\dimexpr \aboverulesep + \belowrulesep + \cmidrulewidth]{fp32} & \multirow{6}{*}[-\dimexpr \aboverulesep + \belowrulesep + \cmidrulewidth]{fp32} & {\color{red}stagnates} \\
& 28 & & & & & {\color{red}stagnates} \\
\cmidrule{1-2} \cmidrule{7-7}
\multirow{2}{*}{crystal} & 8 & & & & & {\color{blue}converges} \\
& 28 & & & & & {\color{red}stagnates} \\
\cmidrule{1-2} \cmidrule{7-7}
\multirow{2}{*}{allreduce} & 8 & \multirow{2}{*}[-\dimexpr \aboverulesep + \belowrulesep + \cmidrulewidth]{fp32+fp64sqrt} & & & & {\color{blue}converges} \\
& 28 & & & & & {\color{blue}converges} \\
\bottomrule
\end{tabular}
\caption{Evaluation of mixed-precision strategies in Nekbone. We show the effect on convergence ($tol=10^{-10}$) of different strategies. We lower precision in the CG loop including the preconditioner. For both, we separate the FP operations and the Gather-Scatter Operations (GSO). The gather-scatter library in Nekbone offers three reduction modes: pairwise (default), crystal, and allreduce.}
\label{tab:evaluation}
\end{table}

\section{Neko case study}
\label{sec:neko}
Neko~\cite{neko} is a portable simulation framework based on high-order spectral elements on hexahedral meshes, mainly focusing on incompressible flow simulations. The framework is written in modern Fortran. It adopts an object-oriented approach, which allows for multi-tier abstractions of the solver stack and facilitates multiple hardware backends, which range from general-purpose processors to accelerators and vector processors. It also includes limited FPGA support. Neko focuses on single core/ single accelerator efficiency via tensor product operator evaluations. 
A key to achieving good performance in spectral element methods is to consider a matrix-free formulation, where one always works with the unassembled matrix on a per-element basis. The Gather–Scatter Operations (GSO) are used to ensure continuity of functions at the element level, operating on both intra-node and inter-node element data.

The primary consideration in Neko is how to efficiently utilize different computer platforms without re-implementing the entire framework for each supported backend. This problem is solved by hiding the implementation of backend-dependent low-level kernels behind a common interface realized as an abstract Fortran type. This way, types that capture higher-level concepts, such as the fluid solver, can remain completely backend-agnostic.

Neko solves the incompressible Navier–Stokes equations in time
\begin{align*}
\frac{\partial u}{\partial t} + (u\cdot\nabla)u &= -\nabla p + \frac{1}{Re}\nabla^2u+f,\\
\nabla\cdot u &= 0,
\end{align*}
where $u$ is the velocity, $p$ is the pressure, $f$ is the volume force and the Reynolds number $Re=UL/v$, with the reference velocity and length $U$ and $L$, and the kinematic viscosity $v$.

At each time step, a Poisson's equation for pressure is solved using the extrapolated velocities on the boundaries. A Helmholtz equation for velocity follows that. The underlying solvers are matrix-free Krylov-type solvers:  Generalized minimal residual method (GMRES) with a block Jacobi preconditioner for velocity, and Conjugate Gradient (CG) with a hybrid additive Schwartz preconditioner for the Poisson's equation. CG with the identity and Jacobi preconditioners is also available. 

\subsection{Code inspection with Verificarlo}
In this study, we are focused on using Neko to solve  Poisson's equation, which is provided as an example within the Neko package, using the CG method with the identity and Jacobi preconditioners. We note the similarity of this problem to the one solved by Nekbone, and hence rely on the previous results in~\Cref{sec:code-inspection}.

\subsection{Enabling mixed-precision in Neko}

We adopt the same strategy explored in Nekbone in which the main code runs in double precision while the solver and its preconditioner operate in single precision (fp64–fp32). \Cref{tab:evaluation-neko} highlights the applied strategies, which we discuss in detail below.

Neko is developed in a highly modular style, and it controls the working precision through a global variable \texttt{rp}, which is set during the configuration stage. Although both double and single precisions are supported, only one can be selected at compile time for the entire program run. 

Hence, we introduced several auxiliary classes that allow precision cropping from fp64 to fp32, both within the code and for global communication operations. This was necessary because, while the software natively supports increased precision via the {\tt xp} mode (used, for instance, in {\tt MPI\_Allreduce}), it does not provide similar support for precision reduction, i.e., switching from fp64 to fp32. Thus, we rewrote some base classes for supporting mixed-precision operations like coefficients ({\tt coef}), function space ({\tt space}), matrix-vector multiplication ({\tt ax}), boundary condition base class ({\tt bc}), Dirichlet boundary condition ({\tt dirichlet}), for storing solution $x$ ({\tt field}), gather-scatter communication ({\tt gs}), Krylov solver base class ({\tt krylov}), and CG solver implementation ({\tt cg}). The idea is to maintain both double and single precision data types within the classes by specifying the precision explicitly. In such a way, the mixed-precision versions of the classes can support type castings for convenient precision switching. We keep the initialization stage in double precision, while the CG loop calculations are in mixed fp64-fp32 precision. For the initialization stage, we let the constant values be set in double precision, to then be cast or copied to single precision for future single precision calculations. This approach introduces only one-time type-castings before the CG loop.

Although this approach performs well overall, we observed that convergence tends to stagnate around $tol=10^{-3}$ (see~\Cref{fig:neko-comparison-five}). 
We attribute this issue to the parallel reductions, and to the global communication implemented via the Gather-Scatter (GS) library similarly to the Nekbone case. GS is an individual module in Neko that allows users to decide the working precision during the configuration.
To enable this mixed-precision communication within PCG, we rewrote the PCG solver interface to accept two GS handlers, one for double-precision and another for single-precision communication. 
Thus, we combine a single precision PCG loop with the global communication via the Gather-Scatter library performed in double precision to ensure both fast and robust convergence in Neko.

\subsubsection{Extended precision for global communication}
\begin{figure}[!ht]
    \centering
    \hspace*{-1mm}\includegraphics[width=1.04\linewidth]{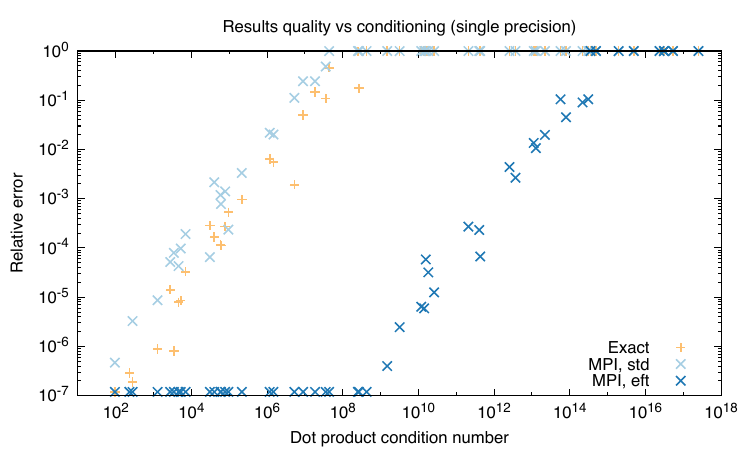}
    \caption{{\tt dot2} results for extending working precision: Exact means sequential dot product, while MPI eft stands for dot2.}
    \label{fig:dot2}
\end{figure}

In practice, due to the casting before and after each dot product (on very large arrays in the preconditioner), it may be costly to enable double-precision dot products within the CG loop while keeping calculations in single precision. Hence, we propose to employ computer arithmetic techniques such as compensated summation and floating-point expansion. While compensated summation (e.g., the Kahan summation) significantly reduces the numerical error~\cite{Hig02}, the floating-point expansion of size two (one digit for the result and one for the error) doubles the storage capacity of single precision, mimicking double precision. For local reductions, we rely on {\tt dot2}~\cite{Ogita05accuratesum} as an example for the compensated summation algorithm. \Cref{fig:dot2} shows the benefit of extended single precision {\tt dot2}: The propagation of the accumulated error is delayed until $cond(xy^T)=10^8$, while the machine precision accuracy $2^{-23}\approx 1.19 \times 10^{-7}$ is preserved. Hence, by using single precision {\tt dot2} we can mimic the double precision dot product without casting. We use {\tt eftdot2} from the libeft library~\footnote{https://github.com/ffevotte/libeft}, which is tuned for Intel CPUs, as well as our generic implementation. The local reduction with {\tt dot2} is followed by the global reduction via the customized MPI operation based on the error-free transformation {\tt twosum}~\cite{Dek71,Knu97} applied to the summation of the floating-point expansion of size two. As we observed, just using {\tt dot2} and the customized MPI allreduce operations bring only marginal improvement to the convergence and performance. This is possibly due to the cached vectors limiting a possibility to hide six extra flops per each element-wise product using dot2; notably, dot2 on large enough vectors (off cache) shows comparable performance to the standard dot product. 

Further investigation is required to get more understanding on this, which we leave to future work. 

To sum up, from~\Cref{tab:evaluation-neko}, we can conclude that the main source of inaccuracy lies in the gather-scatter operations.

For the case with the Jacobi preconditioner (see step 2 in~\Cref{tab:cg-profiling}: $z = J^{-1}r$, where $J^{-1} ~= 1/diag(A)$), we primarily follow the same strategy as above by keeping the initialization in double precision and conducting one-time type castings before the PCG loop to get a single precision copy of the preconditioner. So, we use the single precision preconditionner at the PCG iterations.

\begin{table}[!ht]
\centering
\small
\begin{tabular}{@{}l@{\hspace{4pt}}c@{\hspace{4pt}}c@{\hspace{4pt}}cc@{\hspace{4pt}}cc@{}}
\toprule
\multicolumn{7}{c}{\textbf{Mixed-precision strategy evaluation}} \\
\midrule
\textbf{GSO mode} & \textbf{Init} & {\textbf{Precond.}} & \multicolumn{3}{c}{\textbf{PCG}} & \textbf{Conv.} \\
\cmidrule(lr){3-3} \cmidrule(lr){4-6}
& & \textbf{Ops} & \textbf{Ops} & \textbf{GSO} & \textbf{dot} & \\
\midrule
\multirow{4}{*}{pairwise} & \multirow{4}{*}{fp64} & \multirow{4}{*}{fp32} & \multirow{4}{*}{fp32} & fp32 & fp32 &{\color{red}stagnates} \\
& & & & fp64 & fp64 & {\color{blue}converges} \\
& & & & fp32 & dot2 & {\color{red}stagnates} \\
& & & & fp64 & dot2 & {\color{blue}converges} \\
\bottomrule
\end{tabular}
\caption{Evaluation of mixed-precision strategies in Neko on 28 MPI ranks on Kebnekaise. We show the effect on convergence ($tol=10^{-9}$) of different strategies. We lower precision in the CG loop including the preconditioner. For both, we separate the standard FP operations, dot products, and the Gather-Scatter Operations (GSO), which are only pairwise.}
\label{tab:evaluation-neko}
\end{table}

\section{Results and performance}
\label{sec:results}
We evaluate the performance of the mixed-precision Nekbone and Neko in terms of three dimensions: accuracy, time-to-solution, and energy-to-solution. We set the CG loop stopping criteria to $1.0 \times 10^{-10}$ for Nekbone and $1.0 \times 10^{-9}$ for Neko. We note, however, that in practice, real-world applications may not need such high accuracy, and the convergence may be stopped much earlier. 

The codes are compiled with
\begin{itemize}
\item {\tt flang} version 14.0.1 and Intel {\tt ifort} version 2021.4.0 with the {\tt -O2} optimisation level on the Kebnekaise Skylake nodes equipped with two Intel Xeon Gold 6132 CPUs with 14 cores @2.6~GHz. On Kebnekaise, we also use GNU {\tt gfortran} version 10.2.0 for Neko as flang does not support multi-dimensional arrays.
\item Cray Fortran {\tt ftn} version 16.0.1, and GNU {\tt gfortran} version 11.2.0, with the {\tt -O2} optimisation level on the LUMI-C partition equipped with two AMD EPYC 7763 CPUs with 64 cores @2.45~GHz.
\item Intel {\tt ifort} version 2021.10.0 and the {\tt -O2} optimization level on the MareNostrum 5 (MN5) Accelerated Partition with two Intel Sapphire Rapids 8460Y+ with 40 cores @2.3~Ghz.
\end{itemize}

\subsection{Nekbone}

\subsubsection{Accuracy}
To assess accuracy, we introduce 
absolute error ($AE$) and mean absolute error ($MAE$):\\ 
$\displaystyle
AE=\left| \Delta residual\right| =\left| r_{m} -r_{d} \right|, \hfill \text{and} \hfill MAE=\frac{1}{n}\sum _{i=1}^{n} \left|\Delta residual_{i} \right|,$
where $r_{m}$ is the computed residual for mixed-precision and $r_{d}$ is the computed residual for double precision, and $n$ is the dimension of the residual vectors corresponding to the number of degrees of freedom of the system.

\begin{figure}[!ht]
    \begin{subfigure}{0.49\textwidth}
        \centering
        \includegraphics[width=0.8\linewidth]{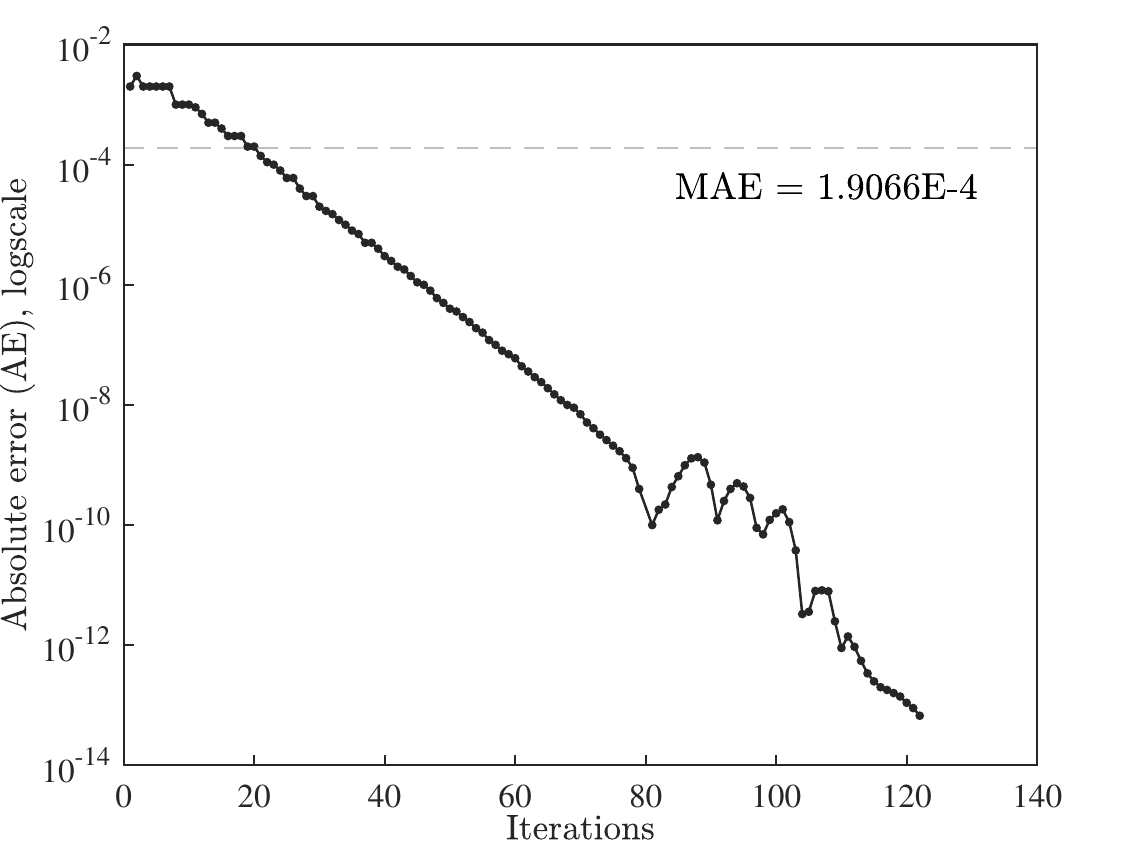}
        \caption{CG without preconditioner} \label{fig:ae-noprecond}
    \end{subfigure}
    \hspace*{\fill}
    \begin{subfigure}{0.49\textwidth}
        \centering
        \includegraphics[width=0.8\linewidth]{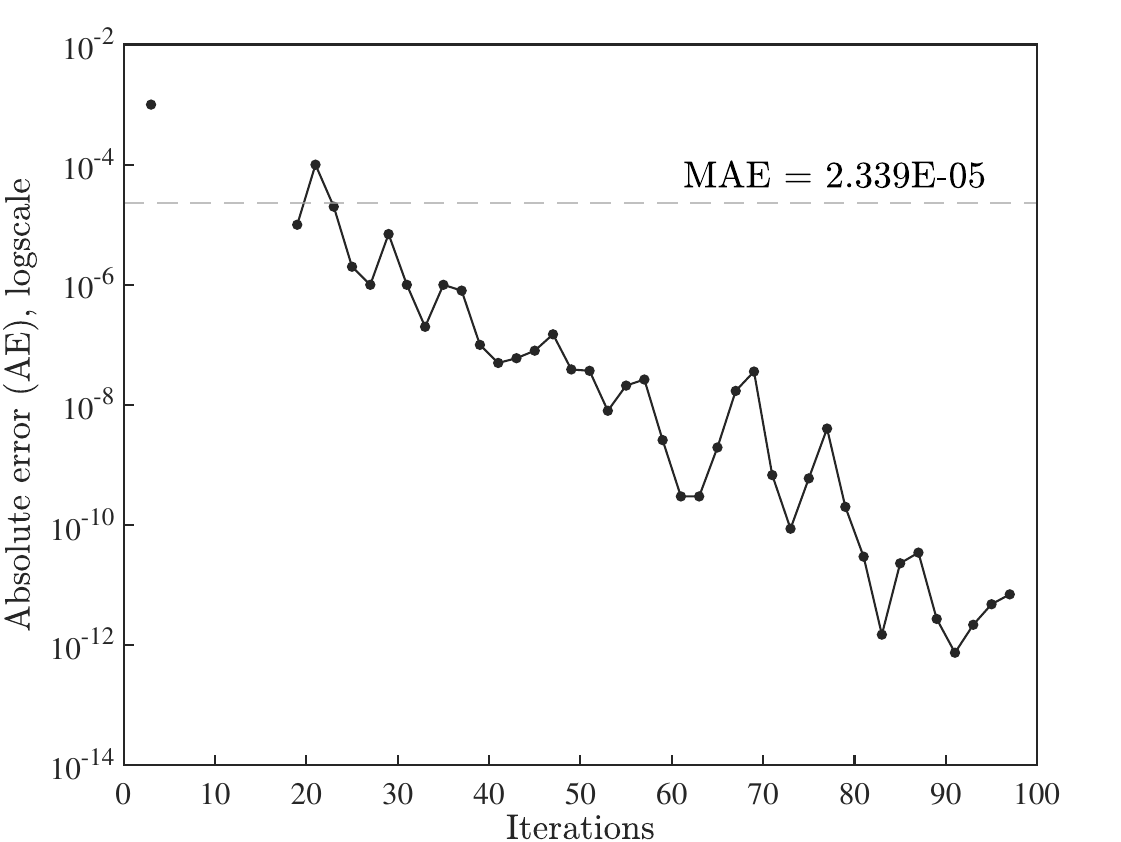}
        \caption{CG with preconditioner} \label{fig:ae-precond}
    \end{subfigure}
    \caption{Absolute error (AE) of the mixed-precision Nekbone against its double version.}
    \label{fig:accuracy-ae}
\end{figure}

\Cref{fig:accuracy-ae} illustrates the history of the absolute error with respect to the number of iterations. For the test case without preconditioner, the error decreases and reaches the tolerance at $iteration=122$ with the value of $6.70 \times 10^{-14}$, with the $MAE$ of $1.91 \times 10^{-4}$. When the preconditioner is enabled, the absolute error decreases with some fluctuations, and the error at convergence ($iteration=97$) is $7.03 \times 10^{-12}$ with the $MAE$ of $2.34 \times 10^{-5}$; for several initial iterations, the error is zero, and is not visible on~\Cref{fig:ae-precond}.

\begin{figure}[!ht]
    \begin{subfigure}{0.49\textwidth}
        \centering
        \includegraphics[width=0.8\linewidth]{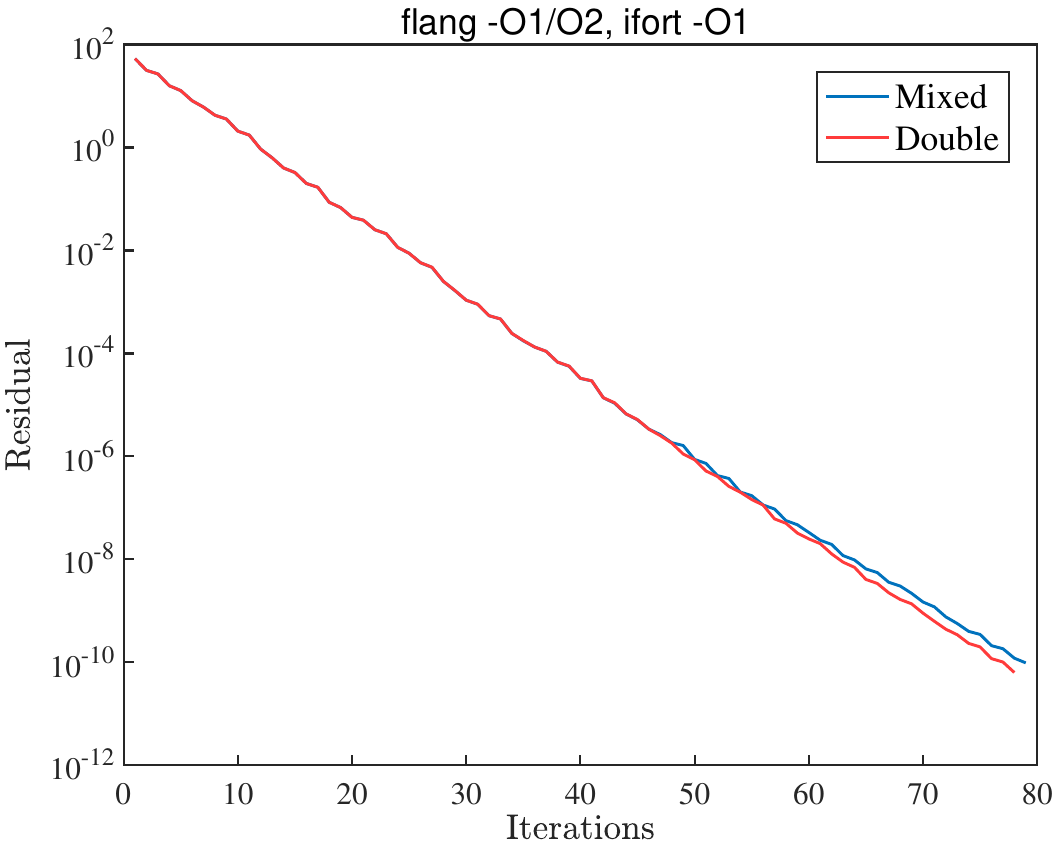}
        \caption{{\tt flang -O1/ -O2}, {\tt ifort -O1}} \label{fig:stagnation-O1}
    \end{subfigure}
    \hspace*{\fill}
    \begin{subfigure}{0.49\textwidth}
        \centering
        \includegraphics[width=0.8\linewidth]{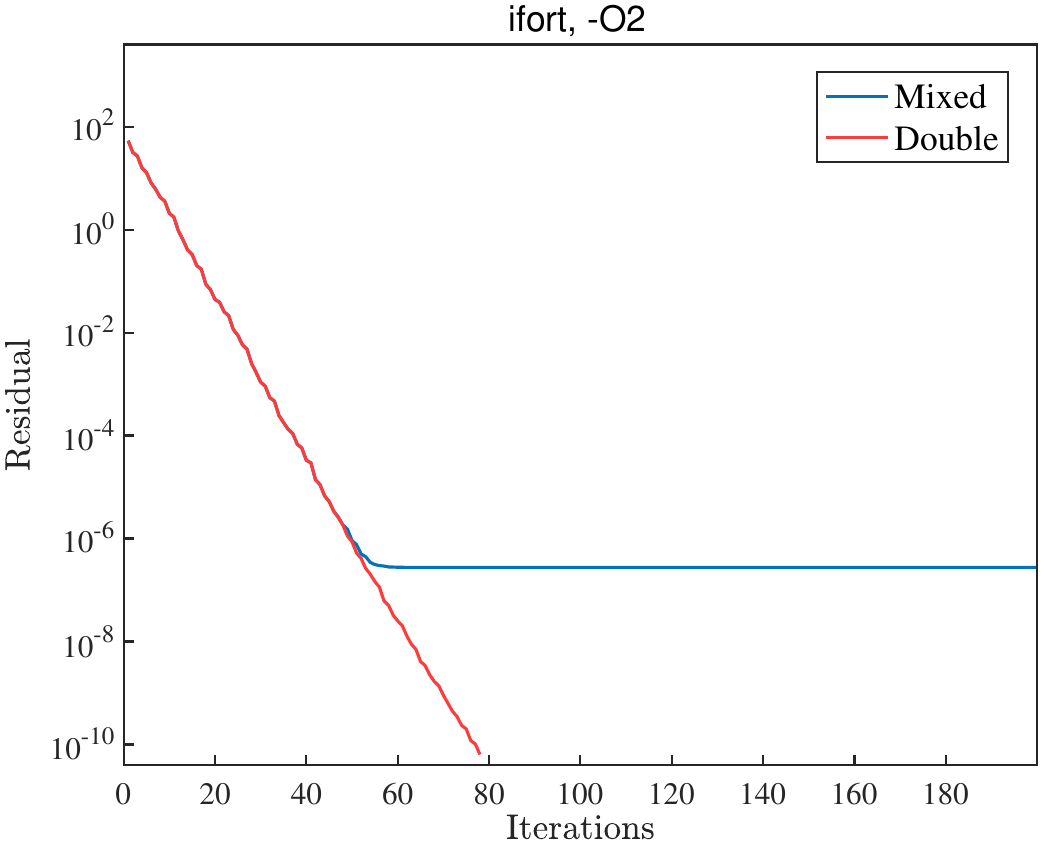}
        \caption{{\tt ifort -O2}} \label{fig:ifort-O2}
    \end{subfigure}
    \caption{Residual history of CG with the multigrid preconditioner in Nekbone
    compiled with {\tt flang \& ifort} under two optimization levels.}
    \label{fig:stagnation-compiler}
\end{figure}

When measuring accuracy of the mixed-precision Nekbone with the preconditioner enabled, we observe stagnation of the residual with {\tt ifort} and the optimisation level {\tt -O2}, see~\Cref{fig:ifort-O2}. This behavior was predicted during the initial inspection phase by the Verificarlo MCA backend, see~\Cref{fig:mca-precond}.  With {\tt flang} and optimisation levels of {\tt -O1} and {\tt -O2}, and with {\tt ifort -O1} there is no stagnation, as depicted in~\Cref{fig:stagnation-O1}; the two plots show sequential runs. However, the stagnation appears even with {\tt flang} when using four MPI ranks as highlighted in~\Cref{tab:evaluation}.


\begin{figure}
    \centering
    \hspace*{-2mm}\includegraphics[width=\linewidth]{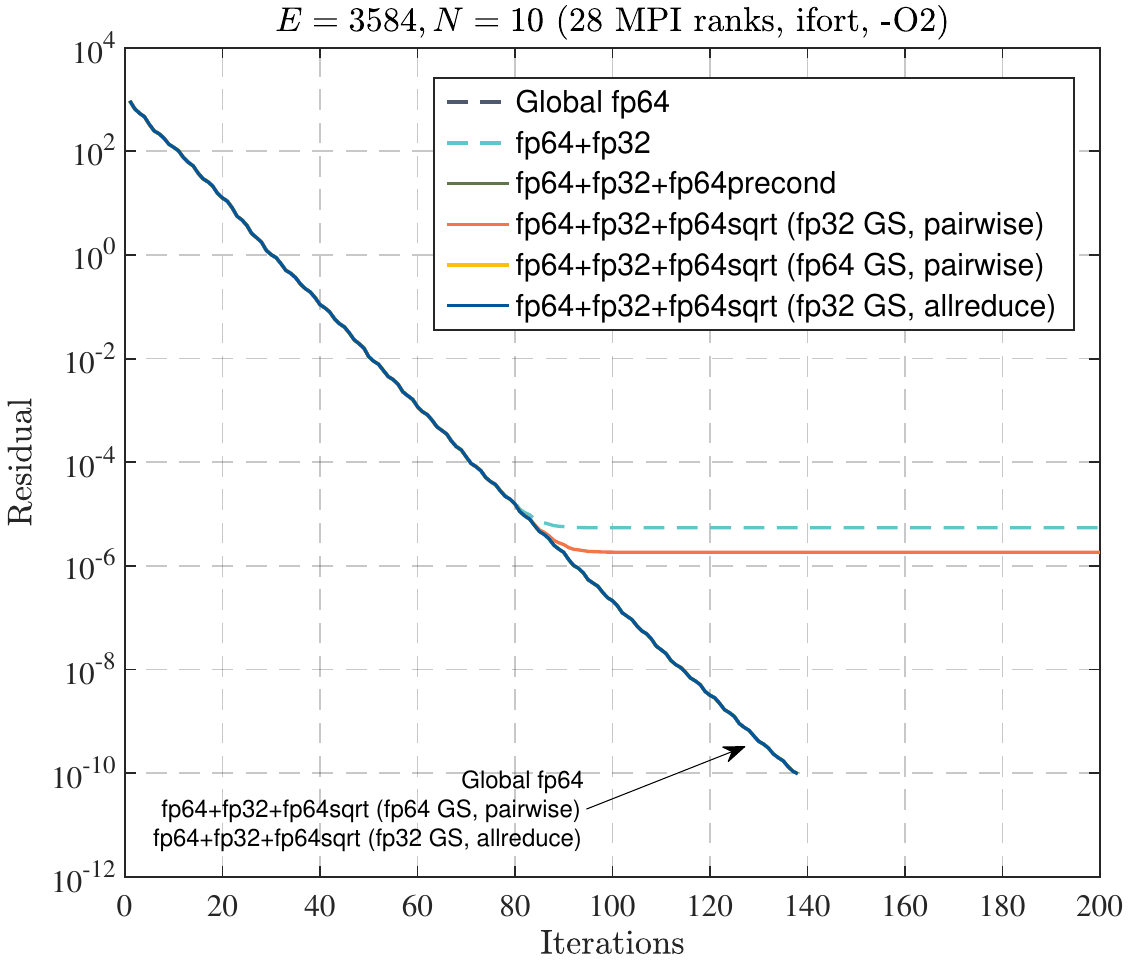}
    \caption{Residual history of CG with the multigrid preconditioner in Nekbone.}
    \label{fig:nekbone-cg-multigrid}
\end{figure}
We implemented the mixed-precision strategy from~\Cref{sec:precond-nekbone}, where the initialization is in double precision, the PCG loop in single, but square roots, dot products, and gather-scatter communication are in double. \Cref{fig:nekbone-cg-multigrid} reports the convergence history of the multigrid preconditioned CG solver within Nekbone, highlighting the success of the proposed strategy. We can see that the pinpointed initial issue in the square root operation only (orange line in~\Cref{fig:nekbone-cg-multigrid}) slightly improves the convergence, and that the biggest need for more accuracy lies in global communication. Such global communication occurs a few times on every iteration of the CG loop as dot products with the corresponding global reductions, but also in the invocation of the gather-scatter operations. All these operations are conducted in double precision with corresponding type castings.

\subsubsection{Time-to-solution}
We conducted tests on both serial (not compiled with MPI) and parallel programs. For the serial tests, we set the number of elements per process to $16$, $50$, 
and $128$, respectively. We measure two types of program runtime: 1) the elapsed time of the whole program, counted with the {\tt time} command in GNU/ Linux; 2) the solve time (the CG loop time), counted by Nekbone internal interfaces that are also included as a part of the output. We report the median of five runs.

\begin{table}[!ht]
    \caption{Elapsed time (secs) of Nekbone, CG without preconditioner on Kebnekaise, {\tt flang}.}    
    \begin{subtable}{.49\linewidth}
      \centering
        \caption{Whole program}
        \begin{tabular}{p{0.22\textwidth}||p{0.17\textwidth}p{0.17\textwidth}p{0.17\textwidth}}
        \hline \noalign{\hrule height 0.3pt}
         Elems & 16 & 50 & 128 \\
        \hline 
         Double & 0.102 & 0.313 & 0.811 \\
         Mixed & 0.096 & 0.273 & 0.669 \\
         \hline
         Gain & 1.06x & 1.15x & 1.21x \\
        \hline \noalign{\hrule height 0.3pt}
        \end{tabular}
    \label{tab:time-serial-noprecond-whole}
    \end{subtable}
    \hspace*{0.2mm}
    \begin{subtable}{.49\linewidth}
      \centering
        \caption{Solve time}
        \begin{tabular}{p{0.17\textwidth}p{0.17\textwidth}p{0.17\textwidth}}

        \hline \noalign{\hrule height 0.3pt}
        16 & 50 & 128 \\
        \hline 
         0.042 & 0.144 & 0.379 \\        
         0.035 & 0.117 & 0.303 \\
         \hline
         1.2x & 1.23x & 1.25x \\
        \hline \noalign{\hrule height 0.3pt}
        \end{tabular}
    \label{tab:time-serial-noprecond-solve}
    \end{subtable}
\label{tab:time-serial-noprecond}
\end{table}
The timing results of the serial program are shown in~\Cref{tab:time-serial-noprecond} for the test case without preconditioner and in~\Cref{tab:time-serial-precond} for the case with preconditioner. The gain is calculated as $Gain=\dfrac{T_{double}}{T_{mixed}}$. 

\begin{figure}[ht]
    \centering
    \includegraphics[width=0.99\linewidth]{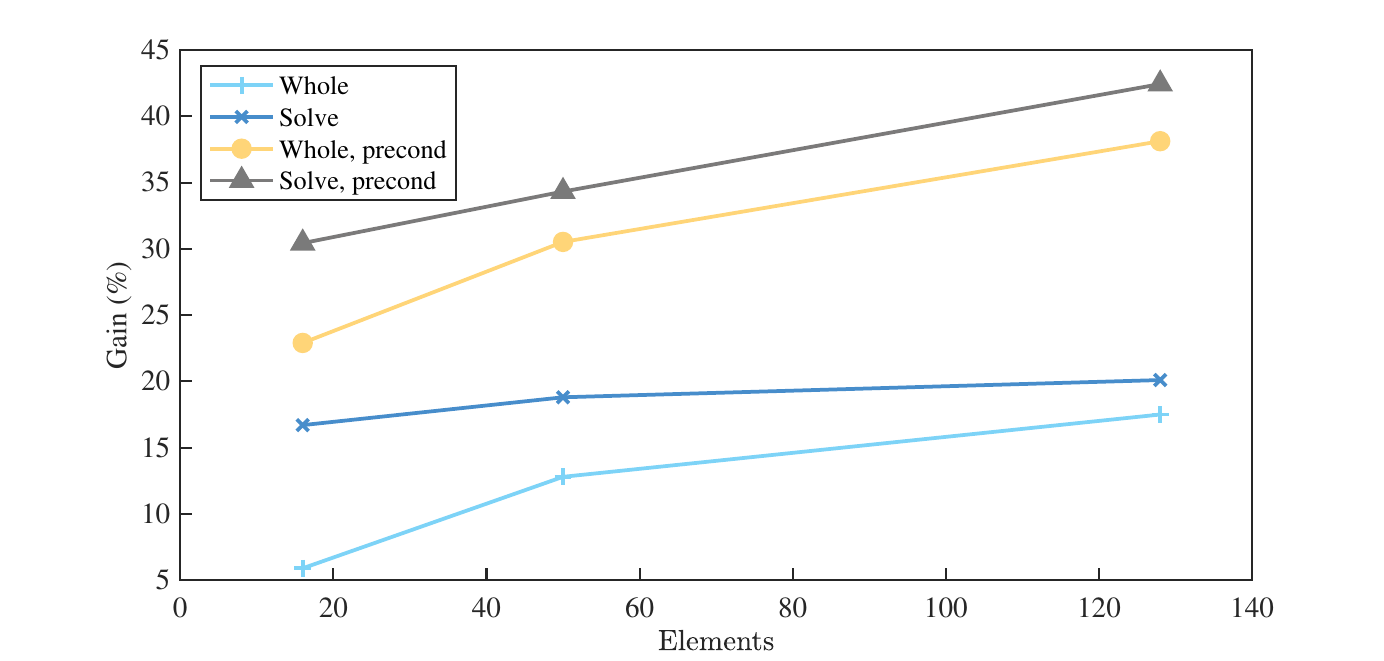}
    \caption{Nekbone, CG without preconditioner: mixed-precision gain in time-to-solution for different elements.}

    \label{fig:time-serial}
\end{figure}

For the case without preconditioner, the gain increases with the number of elements. The mixed-precision version of Nekbone demonstrates significant advantages.
When the preconditioner is enabled, the mixed-precision shows an even greater advantage in both the whole program and the solve time: The gain reaches 1.62x for $128$ elements. \Cref{fig:time-serial} illustrates the benefits of the mixed-precision Nekbone in terms of the reduced time-to-solution for various number of elements for both cases. The gain increases with the number of elements. Notably, the trend is more pronounced in the solve part and for the case with preconditioner, yielding to the 1.74x gain. 
\begin{table}[!ht]
    \caption{Elapsed time (secs) of Nekbone, CG with the multigrid preconditioner on Kebnekaise, {\tt flang}.}
    \begin{subtable}{.49\linewidth}
      \centering
        \caption{Whole program}
        \begin{tabular}{p{0.22\textwidth}||p{0.17\textwidth}p{0.17\textwidth}p{0.17\textwidth}}
        \hline \noalign{\hrule height 0.3pt}
         Elems & 16 & 50 & 128 \\
        \hline 
         Double & 0.214 & 0.806 & 2.634 \\
         Mixed & 0.165 & 0.560 & 1.630 \\         
         \hline
         Gain & 1.3x & 1.44x & 1.62x \\
        \hline \noalign{\hrule height 0.3pt}
        \end{tabular}
    \label{tab:time-serial-precond-whole}
    \end{subtable}%
    \hspace*{1mm}
    \begin{subtable}{.49\linewidth}
      \centering
        \caption{Solve time}
        \begin{tabular}{p{0.17\textwidth}p{0.17\textwidth}p{0.17\textwidth}}
        \hline \noalign{\hrule height 0.3pt}
         16 & 50 & 128 \\
         \hline 
         0.092 & 0.370 & 1.235 \\
         0.064 & 0.243 & 0.711 \\         
         \hline
         1.44x & 1.52x & 1.74x \\
        \hline \noalign{\hrule height 0.3pt}
        \end{tabular}
    \label{tab:time-serial-precond-solve}
    \end{subtable}
\label{tab:time-serial-precond}
\end{table}

For parallel tests without preconditioner, we performed weak scaling tests on one node on LUMI-C with the fixed number of elements $128$ and vary the number of MPI ranks. \Cref{tab:time-mpi-noprecond} reports time-to-solution and highlights the mixed-precision gain. The gain is increasing with the number of processes, resulting in 1.69x 
for the whole program and 2.62x 
for the solve on 128 MPI ranks.
\begin{table}[!ht]
\centering
\caption{Elapsed time (secs) of Nekbone, CG without preconditioner with $128$ elements per MPI rank on LUMI-C, {\tt cray ftn}.}
\begin{subtable}{\linewidth}
    \centering
    \caption{Whole program}
    \begin{tabular}{p{0.165\textwidth}||p{0.065\textwidth}p{0.074\textwidth}p{0.075\textwidth}p{0.08\textwidth}p{0.08\textwidth}p{0.08\textwidth}p{0.08\textwidth}}
    \hline \noalign{\hrule height 0.3pt}
    MPI ranks & 1 & 4 & 8 & 16 & 32 & 64 & 128 \\
    \hline 
    Double & 0.775 & 1.026 & 1.857 & 2.920 & 3.151 & 3.562 & 4.452 \\
    Mixed & 0.741 & 0.905 & 0.995 & 1.642 & 1.694 & 2.096 & 2.641 \\
\hline
Gain & 1.05x & 1.13x & 1.87x & 1.78x & 1.86x & 1.70x & 1.69x \\
    \hline \noalign{\hrule height 0.3pt}
    \end{tabular}
    \label{tab:time-mpi-noprecond-whole}
\end{subtable}
\begin{subtable}{\linewidth}
    \centering
    \caption{Solve time}
    \begin{tabular}{p{0.165\textwidth}||p{0.065\textwidth}p{0.074\textwidth}p{0.075\textwidth}p{0.08\textwidth}p{0.08\textwidth}p{0.08\textwidth}p{0.08\textwidth}}
    \hline \noalign{\hrule height 0.3pt}
    MPI ranks & 1 & 4 & 8 & 16 & 32 & 64 & 128 \\
    \hline 
    Double & 0.182 & 0.239 & 0.596 & 1.115 & 1.161 & 1.207 & 1.245 \\
    Mixed & 0.165 & 0.178 & 0.190 & 0.256 & 0.430 & 0.445 & 0.476 \\
    \hline
    Gain & 1.10x & 1.34x & 3.14x & 4.36x & 2.70x & 2.71x & 2.62x \\
    \hline \noalign{\hrule height 0.3pt}
    \end{tabular}
    \label{tab:time-mpi-noprecond-solve}
\end{subtable}
\label{tab:time-mpi-noprecond}
\end{table}

\subsubsection{Energy-to-solution}
\begin{table}[!ht]
    \centering
    \caption{Energy-to-solution (joules) of Nekbone, CG without preconditioner on LUMI-C.}

    \begin{tabular}{p{0.10\textwidth}||p{0.07\textwidth}p{0.07\textwidth}p{0.07\textwidth}}
    \hline 
    MPI ranks & 32 & 64 & 128 \\
    \hline 
    Double & 990.6 & 1424.8 & 2061.2 \\
    Mixed  & 451.6 & 653.6 & 1089.4 \\    
    \hline
    Gain   & 2.19x & 2.18x & 1.89x \\    
    \hline 
    \end{tabular}
    \label{tab:energy-mpi}
\end{table}
 Measuring the energy consumption of an algorithm is not as simple as measuring the time-to-solution, especially since these type of measurements are not yet widely supported and often require privileged access~\cite{ceec-bpg}. For this, we used a single node on LUMI-C, with exclusive access. To measure the energy-to-solution of the mixed-precision Nekbone, we opted to use the energy accounting plugin provided by Slurm, which is based on {\tt pm\_counters}, due to its ease of use. For more details on energy consumption measurements, we refer to the CEEC Best Practice Guide~\cite{ceec-bpg}.

The energy consumption was retrieved, once the job completed, by {\tt sacct}, using the {\tt --format} option with the desired field {\tt ConsumedEnergy} and specifying the job id. \Cref{tab:energy-mpi} reports the consumed energy in joules of the entire Nekbone with double and mixed-precision versions. We ran each test five times and computed the mean and standard deviation (stddev); we report only the mean values while the stddev was always below 4.5\,\%.  We use the same set up and number of runs as in the performance experiment described in~\Cref{tab:time-mpi-noprecond}. Thus, the reduction in energy-to-solution for the mixed-precision version correlates with the time-to-solution and notably shows better gain, confirming the efficiency of our methodology.

\subsubsection{Performance of the winning strategies}
We highlight the two winning strategies for the mixed-precision Nekbone with the multigrid preconditioner (see also~\Cref{tab:evaluation}):

\begin{itemize}
    \item fp64+fp32+fp64sqrt (fp32 GS, allreduce): initialization in fp64, the PCG loop in fp32, sqrt in the preconditioner in fp64, communication in fp32 with the GSO under the allreduce mode.
    \item fp64+fp32+fp64sqrt (fp64 GS, pairwise): the same as above, but the communication in fp64 and GSO under the pairwise mode.
\end{itemize}
It is worth noting that such a simple operation as the square root is required to be computed in fp64 for convergence of the mixed-precision versions.

\Cref{tab:nekbone-winner-time-kbk,tab:nekbone-winner-time-mn5} show timings of the double precision and two mixed-precision versions on Kebnekaise and MareNostrum 5, respectively. 
The communication strategy has a strong impact on both energy and time-to-solution: generally, pairwise communications behave much better than allreduce mode. Due to that, we clearly distinguish the two communication modes in our double precision runs of Nekbone instead of using the default option, which is determined automatically during the installation. The mixed-precision strategy with the pairwise Gather-Scatter operations outperforms the corresponding double precision version (Double-2) by 1.29x and 1.36x on Kebnekaise and MareNostrum 5, respectively. Similarly. the mixed-precision version under the allreduce mode in GS shows the 1.50x and 1.62x gain compared to the double precision version (Double-1) on Kebnekaise and MareNostrum 5, accordingly.

The energy consumption measurements on MareNostrum5, see~\Cref{tab:nekbone-winner-energy-mn5}, show larger gains compared to the time-to-solution results. With the allreduce communication strategy on 80 MPI ranks, we achieve a 2.43x energy saving. With the pairwise communication strategy, gains are more modest, with a 1.46x improvement. 

\begin{table}[!h]
\caption{Elapsed time (secs) of Nekbone, CG with the multigrid preconditioner, 128 elements per MPI rank on Kebnekaise, {\tt ifort}.}
\centering
\begin{threeparttable}[t]
\begin{tabular}{l||ccccc}
\hline 
MPI ranks & 1 & 4 & 8 & 14 & 28 \\
\hline 
Double-1\tnote{a} & 3.23 & 4.06 & 4.81 & 5.68 & 10.79 \\
Double-2 & 3.26 & 4.11 & 4.75 & 5.17 & 8.67 \\
\cline{2-6} 
Mixed-1\tnote{b} & 2.47 & 3.09 & 3.64 & 4.16 & 7.17 \\
Gain-1 & 1.31x & 1.31x & 1.32x & 1.37x & 1.50x \\
\cline{2-6} 
Mixed-2\tnote{c} & 2.54 & 3.18 & 3.79 & 4.26 & 6.73 \\
Gain-2 & 1.28x & 1.29x & 1.25x & 1.21x & 1.29x \\
\hline
\end{tabular}
\begin{tablenotes}
    \item[a] Double-1: allreduce, Double-2: pairwise
    \item[b] fp64+fp32+fp64sqrt (fp32 GS, allreduce)
    \item[c] fp64+fp32+fp64sqrt (fp64 GS, pairwise)
\end{tablenotes}
\end{threeparttable}
\label{tab:nekbone-winner-time-kbk}
\end{table}

\begin{table}[!h]
\caption{Elapsed time (secs) of Nekbone, CG with the multigrid preconditioner, 128 elements per MPI rank on MareNostrum 5, {\tt ifort}.}
\centering
\begin{threeparttable}[t]
\begin{tabular}{l||cccc}
\hline 
MPI ranks  & 8 & 20 & 40 & 80 \\
\hline 
 Double-1\tnote{a} & 5.79 & 8.98 & 13.33 & 24.02 \\
 Double-2 & 7.55 & 8.62 & 10.84 & 13.71 \\
\cline{2-5} 
 Mixed-1\tnote{b} & 4.79 & 5.99 & 9.42 & 14.85 \\
Gain-1 & 1.21x & 1.50x& 1.42x & 1.62x \\
\cline{2-5}
 Mixed-2\tnote{c} & 4.88 & 6.02 & 8.37 & 10.11 \\
Gain-2 & 1.55x & 1.43x & 1.3x & 1.36x \\
\hline
\end{tabular}
\begin{tablenotes}
    \item[a] Double-1: allreduce, Double-2: pairwise
    \item[b] fp64+fp32+fp64sqrt (fp32 GS, allreduce)
    \item[c] fp64+fp32+fp64sqrt (fp64 GS, pairwise)
\end{tablenotes}
\end{threeparttable}
\label{tab:nekbone-winner-time-mn5}
\end{table}

\begin{table}[!h]
\caption{Energy-to-solution (joules) of Nekbone, CG with the multigrid preconditioner, 128 elements per MPI rank on MareNostrum 5, {\tt ifort}.}
\centering
\begin{threeparttable}[t]
\begin{tabular}{l||cccc}
\hline 
MPI ranks & 8 & 20 & 40 & 80 \\
\hline 
 Double-1\tnote{a} & 1960 & 2865 & 7147 & 18014 \\
 Double-2 & 1877 & 2150 & 4457 & 4920 \\
\cline{2-5} 
 Mixed-1\tnote{b} & 939 & 2088 & 3133 & 7428 \\
 Gain-1 & 2.09x & 1.37x & 2.28x & 2.43x \\
\cline{2-5} 
 Mixed-2\tnote{c} & 934 & 1566 & 2560 & 3367 \\
 Gain-2 & 2.01x & 1.37x & 1.74x & 1.46x \\
\hline
\end{tabular}
\begin{tablenotes}
    \item[a] Double-1: allreduce, Double-2: pairwise
    \item[b] fp64+fp32+fp64sqrt (fp32 GS, allreduce)
    \item[c] fp64+fp32+fp64sqrt (fp64 GS, pairwise)
\end{tablenotes}
\end{threeparttable}
\label{tab:nekbone-winner-energy-mn5}
\end{table}

\subsection{Neko}

\begin{figure}
    \centering
    \includegraphics[width=\linewidth]{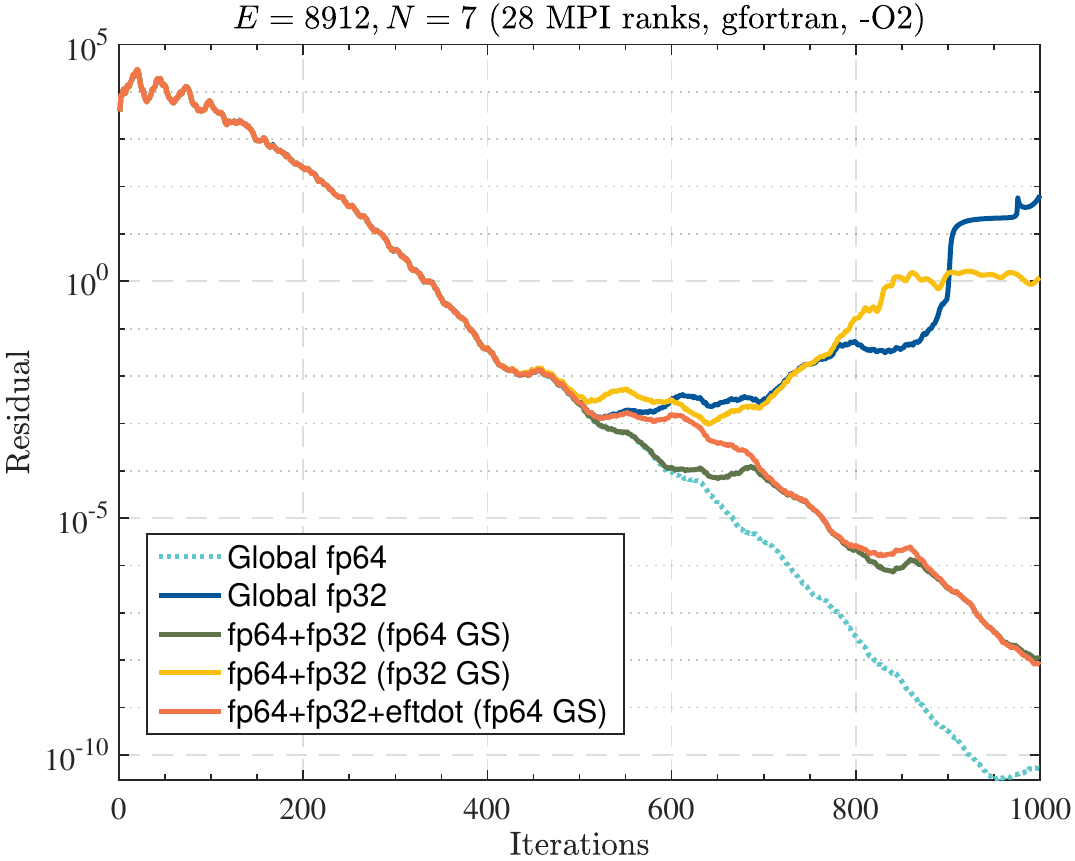}
    \caption{Neko mixed-precision results with dot2 and the Gather-Scatter operations in fp32 or fp64 on Kebnekaise; CG with the identity preconditioner.}
    \label{fig:neko-comparison-five}
\end{figure}

In our experiments, we extended the work of Neko’s team presented in~\cite{neko-reducing-comm}, which focuses on reducing communication overhead in the Conjugate Gradient (CG) method. Their results~\cite[Figure~5]{neko-reducing-comm} highlight a stagnation issue when moving all operations to single precision (fp32) for the Poisson's equation on a cubic domain, with a polynomial degree of $N=7$ and $E=8192$ elements. The CG solver was used with an identity preconditioner. Our observations and analysis confirm this stagnation; see a blue line on~\Cref{fig:neko-comparison-five}. 

To address this stagnation, we repeated the experiment using our proposed strategy, where global operations are conducted in double precision via the Gather-Scatter (GS) library. \Cref{fig:neko-comparison-five} illustrates that switching to single precision for global operations leads to stagnation. However, our mixed-precision strategy -- executing global operations and the initialization in fp64 while keeping the solver and the preconditioner in fp32 -- enhances convergence, allowing the CG solver with the identity preconditioner to reach a tolerance of $1.09 \times 10^{-8}$ on one entire node of Kebnekaise. These global operations occur within the Gather-Scatter library, as well as in dot products, but only for global reductions via {\tt MPI\_Allreduce}. It is worth noting that these GSO are implemented in the pairwise mode only, compared to the three modes in Nekbone (see~\Cref{tab:evaluation}), for the performance benefit at large scale. 

To further improve accuracy, we incorporated the compensated dot product (dot2) locally followed by {\tt MPI\_Allreduce} in double precision, which resulted in a slight improvement, achieving a tolerance of $8.0 \times 10^{-9}$ (see orange line on~\Cref{fig:neko-comparison-five}). Notably, our tests show that extended or higher precision solely for dot products does not significantly improve convergence.  

\begin{figure}
    \centering
    \includegraphics[width=\linewidth]{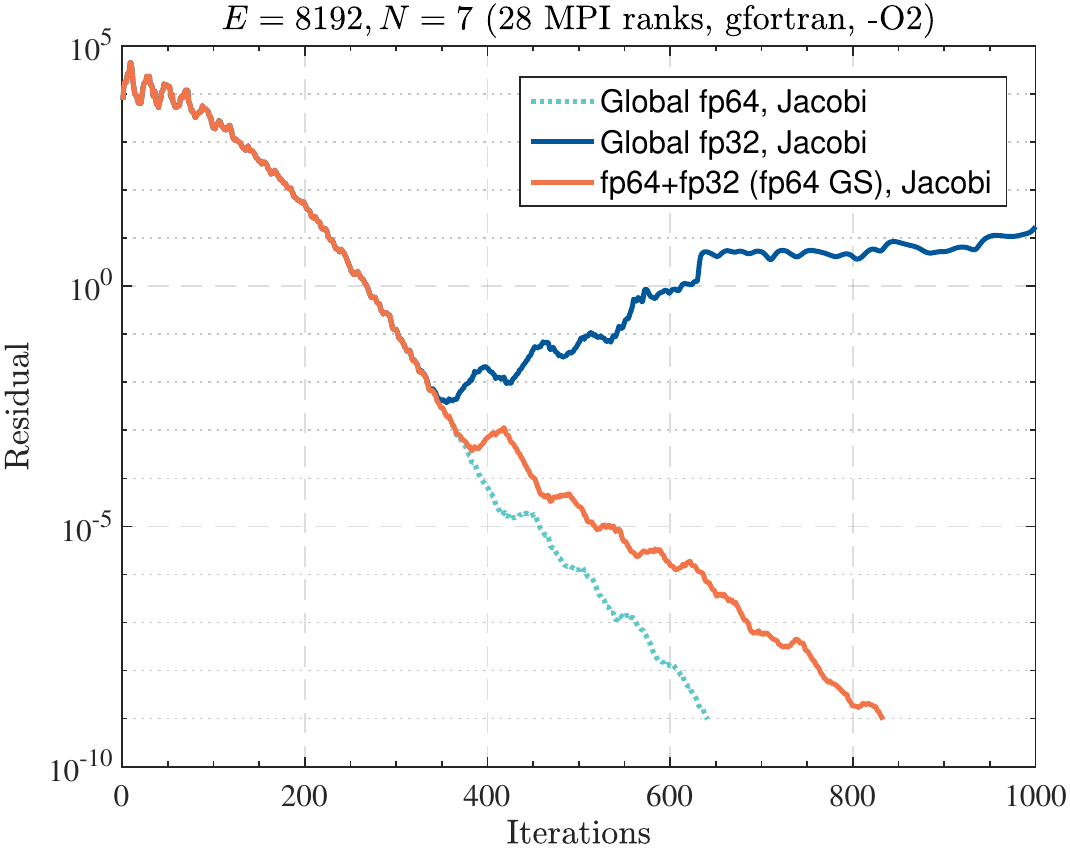}
    \caption{Neko mixed-precision results; CG with the Jacobi preconditioner on Kebnekaise.}
    \label{fig:neko-jacobi}
\end{figure}

We built up and applied our strategy to the CG solver with the Jacobi preconditioner in Neko to solve the same Poisson's example. \Cref{fig:neko-jacobi} shows the convergence history of single and double-precision versions and our mixed-precision version with global communication in double precision. Compared to the case with the identity preconditioner, our mixed-precision implementation converges faster than the full double-precision version. Thus, we also achieved faster convergence in addition to the lighter iterations.

\subsubsection{Time-to-solution}
We evaluated the time-to-solution gain of our mixed-precision strategy for the case with both identity and Jacobi preconditioners on Kebnekaise, LUMI-C, and MareNostrum 5. We used the test case with the number of elements of 16384 and the polynomial degree of 7. 
Since we ran on three different systems, we describe the compiler settings for each of them: On Kebnekaise, we use the gfortran compiler version 10.2.0; on LUMI, we use the gfortran compiler version 11.2.0 as the Cray ftn compiler does not work for current mixed-precision implementation\footnote{The Neko development team also reported issues with this compiler. for details, see the Neko GitHub repository.}. On MareNostrum 5, we rely on the Intel ifort compiler version 2021.10.0. For all settings, we use {\tt -O2} as the compiler optimization option. 

We report the time of the PCG solver as the elapsed time. For the entire application, the mixed-precision Neko always yields a time-to-solution gain of approximately 2 seconds -- between $6.25\%$ and $13.33\%$ faster than the double precision version. Neko's highly modular design currently necessitates additional initializers to support mixed-precision arithmetic, which partially offsets the advantages of mixed-precision at the application level. However, the gain is still clearly noticable in the energy-to-solution metric conducted on the entire application.

For the case with the identity preconditioner, \Cref{tab:neko-identity-time-kbk,tab:neko-identity-time-lumi,tab:neko-identity-time-mn5} report the elapsed time in seconds for the double precision version of Neko as well as for the winning mixed-precision strategy, see fp64-fp32 (fp64 GS) on~\Cref{fig:neko-comparison-five}.
On Kebnekaise, see~\Cref{tab:neko-identity-time-kbk}, we observe some gain on one and four MPI processes. However, the gain is diminished on the larger core counts. We attribute this drawback to the type casting at every iteration of the PCG solver. On LUMI-C, see~\Cref{tab:neko-identity-time-lumi}, our mixed-precision strategy yields up to 1.33x gain in time-to-solution on 64 cores, but on the full node it is slightly slower. On MareNostrum5, see~\Cref{tab:neko-identity-time-mn5}, as on LUMI-C, the best-attained gain for this case is on a half of a node, 1.41x, however the benefit of mixing precision extend to the entire node, yielding to a 1.21x improvement. 
\begin{table}[!h]
\centering
\caption{Elapsed time (secs) of Neko with identity preconditioner on Kebnekaise, {\tt gfortran}.}
\begin{tabular}{l||ccccc}
\hline 
MPI ranks & 1 & 4 & 8 & 14 & 28 \\
\hline 
Double & 245.80 & 68.16 & 37.29 & 23.57 & 11.32 \\
Mixed & 196.60 & 66.56 & 38.09 & 24.41 & 11.80 \\
\hline
Gain & 1.25x & 1.02x & - & - & - \\
\hline
\end{tabular}
\label{tab:neko-identity-time-kbk}
\end{table}
\begin{table}[!h]
\centering
\caption{Elapsed time (secs) of Neko with identity preconditioner on LUMI-C, {\tt gfortran}.}
\begin{tabular}{l||ccccc}
\hline 
MPI ranks & 8 & 16 & 32 & 64 & 128 \\
\hline 
Double & 21.07 & 18.07 & 13.59 & 4.51 & 1.61 \\
Mixed & 23.25 & 17.85 & 11.17 & 3.39 & 1.75 \\
\hline
Gain & - & 1.01x & 1.22x & 1.33x & - \\
\hline
\end{tabular}
\label{tab:neko-identity-time-lumi}
\end{table}
\begin{table}[!h]
\centering
\caption{Elapsed time (secs) of Neko with identity preconditioner on MareNostrum 5, {\tt ifort}.}
\begin{tabular}{l||cccc}
\hline 
MPI ranks & 8 & 20 & 40 & 80 \\
\hline 
Double & 14.10 & 6.61 & 4.80 & 1.87 \\
Mixed & 12.50 & 5.77 & 3.41 & 1.55 \\
\hline
Gain & 1.13x & 1.15x & 1.41x & 1.21x \\
\hline
\end{tabular}
\label{tab:neko-identity-time-mn5}
\end{table}

\begin{table}[!h]
\centering
\caption{Elapsed time (secs) of Neko, CG with the Jacobi preconditioner on Kebnekaise, {\tt gfortran}.}
\begin{tabular}{l||ccccc}
\hline 
MPI ranks & 1 & 4 & 8 & 14 & 28 \\
\hline 
Double & 238.50 & 59.96 & 36.90 & 23.43 & 11.24 \\
Mixed & 170.10 & 57.66 & 25.81 & 19.96 & 9.59 \\
\hline
Gain & 1.4x & 1.04x & 1.43x & 1.17x & 1.17x \\
\hline
\end{tabular}
\label{tab:neko-jacobi-time-kbk}
\end{table}

For the case with the Jacobi preconditioner in Neko for solving Poisson's example, \Cref{tab:neko-jacobi-time-kbk,tab:neko-jacobi-time-lumi,tab:neko-jacobi-time-mn5} demonstrate the time-to-solution of the double and mixed-precision versions. Compared to the results with the identity preconditioner, this test case shows a positive gain of 1.17x on a half and a full node on Kebnekaise. The time-to-solution on LUMI-C is similar to the identity preconditioner case. At the same time, on MareNostrum 5, there is a small decline in the time-to-solution gain, reaching 1.23x and 1.27x on half and full node runs, accordingly.
\begin{table}[!h]
\centering
\caption{Elapsed time (secs) of Neko, CG with the Jacobi preconditioner on LUMI-C, {\tt gfortran}.}
\begin{tabular}{l||ccccc}
\hline 
MPI ranks & 8 & 16 & 32 & 64 & 128 \\
\hline 
Double & 18.83 & 16.50 & 11.52 & 3.57 & 1.37 \\
Mixed & 18.10 & 14.40 & 9.05 & 2.70 & 1.37 \\
\hline
Gain & 1.04x & 1.15x & 1.27x & 1.32x & - \\
\hline
\end{tabular}
\label{tab:neko-jacobi-time-lumi}
\end{table}

\begin{table}[!h]
\centering
\caption{Elapsed time (secs) of Neko, CG with the Jacobi preconditioner on MareNostrum 5, {\tt ifort}.}
\begin{tabular}{l||cccccc}
\hline 
MPI ranks & 8 & 20 & 40 & 80 \\
\hline 
Double & 11.12 & 5.46 & 3.75 & 1.71 \\
Mixed & 10.50 & 5.16 & 3.06 & 1.35 \\
\hline
Gain & 1.06x & 1.06x & 1.23x & 1.27x \\
\hline
\end{tabular}
\label{tab:neko-jacobi-time-mn5}
\end{table}

\subsubsection{Energy-to-solution}
We conducted energy measurements on LUMI-C and MareNostrum 5 with exclusive access in order to obtain clean, undisturbed data. While on LUMI-C we relied on the slurm {\tt sacct} pluggin for measuring energy consumption, on MareNostrum 5 we used the Energy Aware Runtime (EAR)~\cite{earpaper}, which is a collection of tools offering system power consumption and job energy accounting. To measure energy consumption with EAR, we added a few lines in the slurm job batch script (e.g., enable EAR and specify its monitoring policy~\cite{earbsc}). Then, after the job completion, we ran the command {\tt eacct} (similar to {\tt sacct}) to obtain energy consumption. As for the time-to-solution, we run each job at least five times and compute mean.

\Cref{tab:neko-identity-energy-lumi,tab:neko-identity-energy-mn5,tab:neko-jacobi-energy-lumi,tab:neko-jacobi-energy-mn5} report the results of harvesting energy consumption of double precision and our winning mixed-precision strategy on both LUMI-C and MareNostrum 5. Here, we focus on larger core counts, at least half of the chip. We note that these results are coherent with the time-to-solution measurements in~\Cref{tab:neko-identity-time-lumi,tab:neko-identity-time-mn5,tab:neko-jacobi-time-lumi,tab:neko-jacobi-time-mn5}, for the same problem size and the polynomial degree, in terms of benefits from mixing precisions. On both clusters, the energy-to-solution gain is lower than the time-to-solution gain. We attribute this to more optimized code of a real-world application such as Neko and the frequent type casting. On LUMI-C, for both the identity and Jacobi preconditioners, the saving is  1.26x on half of a node. On the entire node, the saving of the mixed-precision is 1.06x for the identity and 1.20x for the Jacobi precondition, although there is no gain in terms of time-to-solution. This confirms our claim that the time-to-solution and energy-to-solution do not correlate 1:1. On MareNostrum 5, the saving from the mixed-precision is constantly present with 1.32x for the identity and 1.16x for the Jacobi preconditioner. 

\begin{table}[!h]
\centering
\caption{Energy-to-solution (joules) of Neko, CG with the identity preconditioner on LUMI-C, {\tt gfortran}.}
\begin{tabular}{l||ccc}
\hline 
MPI ranks & 32 & 64 & 128 \\
\hline 
Double & 10302 & 4996 & 2962 \\
Mixed & 9170 & 3960 & 2802 \\
\hline
Gain & 1.12x & 1.26x & 1.06x \\
\hline
\end{tabular}
\label{tab:neko-identity-energy-lumi}
\end{table}

\begin{table}[!h]
\centering
\caption{Energy-to-solution (joules) of Neko, CG with the Jacobi preconditioner on LUMI-C, {\tt gfortran}.}
\begin{tabular}{l||ccc}
\hline 
MPI ranks & 32 & 64 & 128 \\
\hline 
Double & 9113 & 4118 & 2748 \\
Mixed & 7462 & 3268 & 2295 \\
\hline
Gain & 1.22x & 1.26x & 1.20x \\
\hline
\end{tabular}
\label{tab:neko-jacobi-energy-lumi}
\end{table}

\begin{table}[!h]
\centering
\caption{Energy-to-solution (joules) of Neko, CG with the identity preconditioner on MareNostrum 5, {\tt ifort}.}
\begin{tabular}{l||ccc}
\hline 
MPI ranks & 20 & 40 & 80 \\
\hline 
Double & 14437 & 11686 & 10612 \\
Mixed & 13403 & 10509 & 8033 \\
\hline
Gain & 1.08x & 1.11x & 1.32x \\
\hline
\end{tabular}
\label{tab:neko-identity-energy-mn5}
\end{table}

\begin{table}[!h]
\centering
\caption{Energy-to-solution (joules) of Neko, CG with the Jacobi preconditioner on MareNostrum 5, {\tt ifort}.}
\begin{tabular}{l||ccc}
\hline 
MPI ranks & 20 & 40 & 80 \\
\hline 
Double & 11666 & 9598 & 10056 \\
Mixed & 11350 & 9089 & 8669 \\
\hline
Gain & 1.03x & 1.06x & 1.16x \\
\hline
\end{tabular}
\label{tab:neko-jacobi-energy-mn5}
\end{table}

\begin{figure*}
    \centering
    \includegraphics[width=0.86\linewidth]{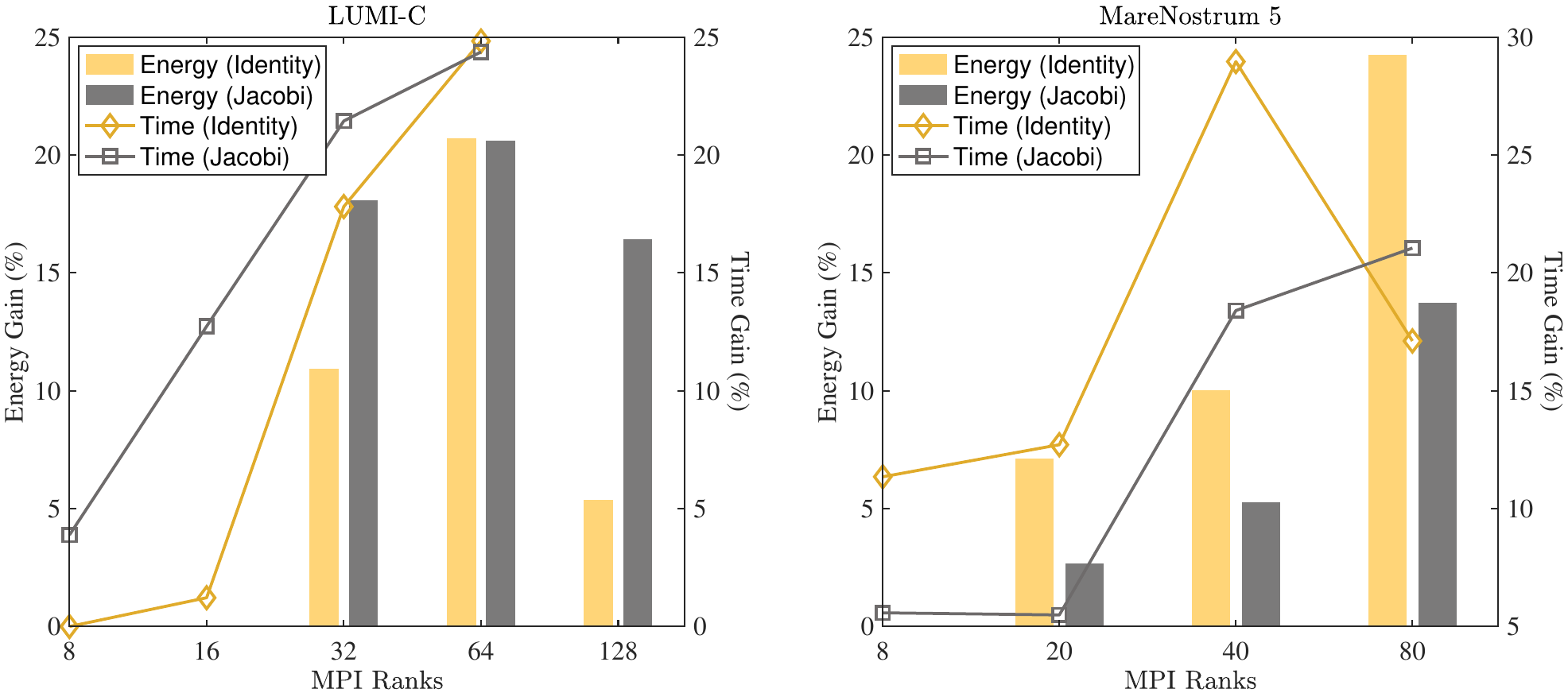}
    \caption{Energy-to-solution gain (Bar, left y-axis) and time-to-solution gain (Curve, right y-axis) of the mixed-precision Neko on LUMI-C and MareNostrum 5.}
    \label{fig:neko-result-bar}
\end{figure*}
\Cref{fig:neko-result-bar} summarizes the benefits of adopting mixed-preci\-sion in Neko for both identity and Jacobi preconditioner in terms of time-to-solution (lines and right y-axis) and energy-to-solution (bars and left y-axis). Notably, on LUMI-C, the left plot, there is no reduction in the mixed-precision execution time, however energy-to-solution is still lower than for the double precision Neko.

\section{Conclusion and Future Work}

\label{sec:conclusion}
We have presented a systematic methodology for enabling mixed-precision in CFD codes by combining detailed code inspection with advanced tools, targeted implementation, and rigorous validation. Using two representative case studies -- Nekbone and Neko -- we demonstrate that carefully applied mixed-precision strategies can yield significant performance improvements while maintaining the required accuracy.

In Nekbone, our mixed-precision implementation reduced time-to-solution on MareNostrum by up to 1.62x and energy-to-solution by 2.43x on 80 MPI ranks by using single precision within the Conjugate Gradient (CG) solver while retaining double precision for critical operations such as global communication and preconditioning. Extending this approach to Neko, we observed robust convergence improvements and notable time-to-solution and energy-to-solution gains (up to approximately 1.3x in select configurations) across different preconditioners. 
Our results confirm that maintaining double precision in global reductions -- via enhanced Gather-Scatter operations -- and for specific operations (e.g., square root and dot products) is important to ensure stable convergence. 

Looking forward, we plan to explore adaptive precision techniques that dynamically adjust the solver’s precision based on convergence behavior, further reducing computational cost while preserving solution quality. In addition, our ongoing work aims to extend these mixed-precision strategies to more larger applications, such as SOD2D from the CEEC project.

Overall, our findings underscore the practical benefits and feasibility of mixed-precision computing in high-performance and high fidelity computational fluid dynamics spectral element applications, bridging legacy codes and modern implementations to achieve substantial improvements in both performance and energy efficiency. The methodology is applicable to other CFD codes and beyond that use the Krylov methods such as CG, BiCGStab, and GMRES as their inner solvers.

Our mixed-precision versions of Nekbone and Neko are available here: \url{https://github.com/yanxchen/enabling-mxp}.
\section*{Acknowledgment}
We thank G\"ulçin Gedik for her help in measuring energy consumption on LUMI-C. This research was partially supported by a Center of Excellence in Exascale CFD (CEEC) grant No 101093393 
funded by the European Union via the European High Performance Computing Joint Undertaking (EuroHPC JU) and Sweden, Germany, Spain, Greece and Denmark.

We acknowledge NAISS in Sweden, partially funded by VR no. 2022-06725, for access to LUMI, EuroHPC JU, and hosted by CSC in Finland. Furthermore, this research was conducted using the resources of HPC2N at Ume\aa{} University, Sweden and MareNostrum5 at BSC, Spain via the EuroHPC Development Access Call.


\begin{thebibliography}{10}
\expandafter\ifx\csname url\endcsname\relax
  \def\url#1{\texttt{#1}}\fi
\expandafter\ifx\csname urlprefix\endcsname\relax\def\urlprefix{URL }\fi
\expandafter\ifx\csname href\endcsname\relax
  \def\href#1#2{#2} \def\path#1{#1}\fi

\bibitem{ppam-paper}
Y.~Chen, P.~de~Oliveira~Castro, P.~Bientinesi, R.~Iakymchuk, Enabling mixed-precision with the help of tools: A Nekbone case study, in: Parallel Processing and Applied Mathematics, PPAM 2024. \href{https://doi.org/10.1007/978-3-031-85697-6_3}{\path{doi:10.1007/978-3-031-85697-6_3}}.

\bibitem{Saa03}
Y.~Saad, Iterative methods for sparse linear systems, 3rd Edition, SIAM, 2003.
\newblock \href{https://doi.org/10.1137/1.9780898718003} {\path{doi:10.1137/1.9780898718003}}.

\bibitem{nekbone}
{Nek5000 developers}, Nekbone, \url{https://github.com/Nek5000/Nekbone}.

\bibitem{Denis2016verificarlo}
C.~Denis, P.~de~Oliveira~Castro, E.~Petit, Verificarlo: Checking floating point accuracy through monte carlo arithmetic, in: ARITH, 2016, pp. 55--62.
\newblock \href {https://doi.org/10.1109/ARITH.2016.31} {\path{doi:10.1109/ARITH.2016.31}}.

\bibitem{parker97}
D.~S. Parker, \href{http://www.cs.ucla.edu/~stott/mca/CSD-970002.ps.gz}{{Monte Carlo} arithmetic: exploiting randomness in floating-point arithmetic}, Tech. Rep. CSD 970002, CS Dept, University of California (1997).
\newline\urlprefix\url{http://www.cs.ucla.edu/~stott/mca/CSD-970002.ps.gz}

\bibitem{IEEE7542019}
{{IEEE}~Computer~Society}, {IEEE} Standard for Floating-Point Arithmetic, {IEEE} Standard 754-2019, 2019.

\bibitem{verificarloproject}
P.~de~Oliveira~Castro, Y.~Chatelain, E.~P. et~al., \href{https://doi.org/10.5281/zenodo.8301008}{verificarlo/verificarlo: Verificarlo v1.0.0} (Aug. 2023).
\newblock \href {https://doi.org/10.5281/zenodo.8301008} {\path{doi:10.5281/zenodo.8301008}}.
\newline\urlprefix\url{https://doi.org/10.5281/zenodo.8301008}

\bibitem{Chatelain2019automatic}
Y.~Chatelain, E.~Petit, P.~de~Oliveira~Castro, G.~Lartigue, D.~Defour, Automatic exploration of reduced floating-point representations in iterative methods, in: Euro-Par, LNCS, Springer, 2019.
\newblock \href {https://doi.org/10.1007/978-3-030-29400-7_34} {\path{doi:10.1007/978-3-030-29400-7_34}}.

\bibitem{nek5000}
P.~F. Fischer, J.~W. Lottes, S.~G. Kerkemeier, nek5000 web page, \url{https://nek5000.mcs.anl.gov}.

\bibitem{gslib}
{Nek5000 developers}, Gather-scatter library, \url{https://github.com/Nek5000/gslib}, accessed 30-JAN-2024.

\bibitem{callgrind}
{Valgrind™ Developers}, Callgrind homepage, \url{https://valgrind.org/docs/manual/cl-manual.html}.

\bibitem{valgrind}
N.~Nethercote, J.~Seward, Valgrind: A framework for heavyweight dynamic binary instrumentation, SIGPLAN Not. 42~(6) (2007) 89--100.
\newblock \href {https://doi.org/10.1145/1273442.1250746} {\path{doi:10.1145/1273442.1250746}}.

\bibitem{iakymchuk19jcam}
R.~{ Iakymchuk}, M.~Barreda, M.~Wiesenberger, J.~I. Aliaga, E.~S. Quintana-Orti, {Reproducibility Strategies for Parallel Preconditioned Conjugate Gradient}, JCAM 371 (2020) 112697, available online 2 January 2020.
\newblock \href {https://doi.org/10.1016/j.cam.2019.112697} {\path{doi:10.1016/j.cam.2019.112697}}.

\bibitem{roofline}
S.~Williams, A.~Waterman, D.~Patterson, Roofline: an insightful visual performance model for multicore architectures, Commun. ACM (2009) 65–76\href {https://doi.org/10.1145/1498765.1498785} {\path{doi:10.1145/1498765.1498785}}.

\bibitem{advisor}
{Intel Corporation}, Intel advisor, \url{https://www.intel.com/content/www/us/en/developer/tools/oneapi/advisor.html}.

\bibitem{neko}
N.~Jansson, M.~Karp, A.~Podobas, S.~Markidis, P.~Schlatter, Neko: A modern, portable, and scalable framework for high-fidelity computational fluid dynamics, Computers \& Fluids 275 (2024) 106243.
\newblock \href {https://doi.org/10.1016/j.compfluid.2024.106243} {\path{doi:10.1016/j.compfluid.2024.106243}}.

\bibitem{Hig02}
N.~J. Higham, Accuracy and stability of numerical algorithms, 2nd Edition, SIAM, 2002.
\newblock \href {https://doi.org/10.1137/1.9780898718027} {\path{doi:10.1137/1.9780898718027}}.

\bibitem{Ogita05accuratesum}
T.~Ogita, S.~M. Rump, S.~Oishi, Accurate sum and dot product, SIAM J. Sci. Comput 26~(6) (2005) 1955--1988.
\newblock \href {https://doi.org/10.1137/030601818} {\path{doi:10.1137/030601818}}.

\bibitem{Dek71}
T.~J. Dekker, A floating point technique for extending the available precision, Numerische Mathematik 18~(3) (1971) 224--242.
\newblock \href {https://doi.org/10.1007/BF01397083} {\path{doi:10.1007/BF01397083}}.

\bibitem{Knu97}
D.~E. Knuth, The Art of Computer Programming: Seminumerical Algorithms, 3rd ed., Vol.~2, Addison-Wesley, 1997.

\bibitem{ceec-bpg}
R.~Iakymchuk, G.~Gedik, K.~Kulkarni, Y.~Chen, D.~Kempf, S.~Kemmler, D.~Papageorgiou, D.~Konioris, S.~Kiebdaj, J.~Corbalan, H.~Köstler, {Best Practice Guide -- Harvesting energy consumption on European HPC systems: Sharing Experience from the CEEC project} (Aug. 2024).
\newblock \href {https://doi.org/10.5281/zenodo.13306639} {\path{doi:10.5281/zenodo.13306639}}.

\bibitem{ceecbpg}
R.~Iakymchuk, G.~Gedik, K.~Kulkarni, Y.~Chen, D.~Kempf, S.~Kemmler, D.~Papageorgiou, D.~Konioris, S.~Kiebdaj, J.~Corbalan, H.~Köstler, Best practice guide--harvesting energy consumption on european hpc systems: sharing experience from the ceec project (Aug. 2024).
\newblock \href {https://doi.org/10.5281/zenodo.13306639} {\path{doi:10.5281/zenodo.13306639}}.

\bibitem{neko-reducing-comm}
M.~Karp, N.~Jansson, A.~Podobas, P.~Schlatter, S.~Markidis, Reducing communication in the conjugate gradient method: a case study on high-order finite elements, in: Proceedings of the Platform for Advanced Scientific Computing Conference, PASC '22, Association for Computing Machinery, New York, NY, USA, 2022.
\newblock \href {https://doi.org/10.1145/3539781.3539785} {\path{doi:10.1145/3539781.3539785}}.

\bibitem{earpaper}
J.~Corbal{\'{a}}n, L.~Alonso, J.~Aneas, L.~Brochard, Energy optimization and analysis with {EAR}, in: {IEEE} International Conference on Cluster Computing, {CLUSTER} 2020, Kobe, Japan, September 14-17, 2020, {IEEE}, 2020, pp. 464--472.
\newblock \href {https://doi.org/10.1109/CLUSTER49012.2020.00067} {\path{doi:10.1109/CLUSTER49012.2020.00067}}.

\bibitem{earbsc}
J.~Corbal{\'{a}}n, B.~Czaja, M.~Kruiter, Tutotial at {ISC} 2024 -- energy management and optimization with ear, details on \url{https://github.com/sara-nl/ISC-2024-EAR-tutorial}.

\bibitem{Deville-Fischer-Mund-2002} Deville, M.O., Fischer, P.F., Mund, E.H. (2002). High-Order Methods for Incompressible Fluid Flow. Cambridge University Press. \newblock \href {https://doi.org/10.1017/CBO9780511546792} {\path{doi:10.1017/CBO9780511546792}}

\end{thebibliography}

\end{document}